\documentclass[aps,reprint,nofootinbib,superscriptaddress]{revtex4}

\usepackage[english]{babel}
\usepackage[a4paper,top=2cm,bottom=2cm,left=3cm,right=3cm,marginparwidth=1.75cm]{geometry}

\usepackage{graphicx}
\usepackage{amssymb}
\usepackage{amsmath}
\usepackage{orcidlink}
\usepackage{bm}
\usepackage{hyperref}
\usepackage{datetime}
\usepackage{xcolor}
\usepackage[T1]{fontenc} 
\usepackage{array}
\usepackage{mathtools}
\usepackage{booktabs}
\usepackage{multirow}
\newcolumntype{C}[1]{>{\centering\arraybackslash}p{#1}} 
\newcommand{\qT}{\boldsymbol{q}_{T}}
\newcommand{\bT}{\boldsymbol{b}_T}
\newcommand{\kperp}{\boldsymbol{k}_\perp}
\newcommand{\bb}{b}
\newcommand{\bmin}{\bb_{\rm min}}
\newcommand{\bmax}{\bb_{\rm max}}
\newcommand{\bstar}{\bb_*}
\newcommand{\mustar}{\mu_{\bstar}}
\newcommand{\ndat}{N_{\rm dat}}

\newcommand{\fourier}[1]{\widetilde{#1}}
\newcommand{\bcut}{\boldsymbol{b}_T^{\text{\footnotesize cut}}}

\begin{document}

\title{The impact of prescriptions in phenomenological extractions of\\Transverse Momentum Dependent distributions}

\author{Matteo Cerutti}
\email{matteo.cerutt@cea.fr \\ ORCID: 0000-0001-7238-5657}
\affiliation{IRFU, CEA, Université Paris-Saclay, F-91191 Gif-sur-Yvette, France}

\author{Andrea Simonelli}
\email{andrea.simonelli@roma1.infn.it \\ ORCID: 0000-0003-2607-9004}
\affiliation{INFN Sezione di Roma, Piazzale Aldo Moro 5, 00185 Roma, Italy}

\begin{abstract}

We investigate the impact of phenomenological prescriptions in the Collins–Soper–Sterman (CSS) approach for global extractions of Transverse Momentum Distributions (TMDs). We show that fits to low-energy Drell-Yan data with different choices of $b_*$ prescription yield equally good agreement with data and similar TMDs at small partonic transverse momentum. In contrast, sizable differences emerge at intermediate transverse momentum region, significantly affecting the predictions for high-energy Drell-Yan processes.
Our results demonstrate that the $b_*$ prescription represents an intrinsic source of theoretical uncertainty in the CSS approach, introducing systematic effects that influence TMD extractions and their interpretation. 
At the same time, our analysis emphasizes the interplay between data at different energy scales in assessing the effect of phenomenological prescriptions in TMD fits adopting the CSS framework. 
\end{abstract}

\maketitle
\tableofcontents

\section{Introduction}
\label{introduction}

The three-dimensional structure of hadrons, the bound states of Quantum Chromodynamics (QCD), provides essential insight into color confinement and hadronization. A powerful tool to explore it is given by transverse-momentum-dependent (TMD) parton densities, which describe how the spin and three-dimensional motion of partons correlate with the spin–momentum properties of the hadron that confines them~\cite{Rogers:2015sqa,Diehl:2015uka,Angeles-Martinez:2015sea,Bacchetta:2016ccz,Scimemi:2019mlf,Boussarie:2023izj,Lorce:2025aqp}. These densities $F_{j/h}$ extend the concept of standard parton distribution functions (PDFs), and are often given an intuitive interpretation as describing the distribution of a parton of species $j$ with light-cone momentum fraction $x$ and transverse momentum $\kperp$ inside a hadron $h$. They arise in factorization theorems of processes sensitive to the transverse momentum $\qT$ of a final state relative to a specified axis, in the kinematic limit where $|\qT|$ is much smaller than the hard scale $Q$ characterizing the observable~\cite{Ji:2004wu,Collins:2011zzd,Echevarria:2011epo}. A notable example is Drell-Yan (DY) lepton-pair production from hadron-hadron collisions, in which the hard scale is given by the invariant mass squared $Q^2$ of the lepton pair and $\qT$ is the transverse momentum of the pair measured with respect to the beam axis. In the kinematic domain of TMD factorization, where $|\qT| \ll Q$, the differential DY cross section can be written as\footnote{Note that we neglected kinematic power corrections~\cite{Vladimirov:2023aot,Piloneta:2024aac}.} 
\begin{equation}
\begin{split}
\label{e:DYZ_xsec}
&\frac{d\sigma^{\text{DY}}}{d |\qT|\, dy\, dQ} = \frac{16 \pi^2 \alpha^2 |\qT|}{9 Q^3}\, x_1\, x_2 \, {\cal H}^{\text{DY}}\big(a_S(\mu);{Q}/{\mu}\big) \, \sum_q c_q(Q^2)
\\
& \qquad \times \int d^2 {\kperp}_1\, d^2 {\kperp}_2\, F_{q/h_1}(x_1,{\kperp^2}_1;\mu,\zeta_1)\, F_{\bar{q}/h_2}(x_2,{\kperp^2}_2;\mu,\zeta_2)\, \delta^{(2)}({\kperp}_1 + {\kperp}_2 - \qT) \; + (1\leftrightarrow2) \; ,
\end{split}
\end{equation}
where $\alpha$ denotes the electromagnetic coupling 
and $x_{1}=Qe^{y}/\sqrt{s}$, $x_{2}=Qe^{-y}/\sqrt{s}$ are the longitudinal momentum fractions carried by the incoming partons, with $y$ the pseudorapidity of the lepton pair. 
For simplicity, we restrict to unpolarized quarks inside unpolarized hadrons $h_1$ and $h_2$. 
The hard function ${\cal H}^{\text{DY}}$ encodes the virtual part of the scattering and depends on $Q$ and on the renormalization scale $\mu$.
It is fully under perturbative control and can be expanded in powers of $a_s=\alpha_s/(4\pi)$~\cite{Collins:2017oxh}. 
The sum runs over all active quark flavors, and $c_{q}$ are the standard electroweak charges (see, \textit{e.g.}, Ref.~\cite{Bacchetta:2019sam} for their specific expressions). 

The second line of Eq.~(\ref{e:DYZ_xsec}) contains the convolution of the unpolarized TMD PDFs $F_{q/h_1}$ and $F_{\bar{q}/h_2}$. They depend on the longitudinal $x_{1,2}$ and transverse ${\kperp}_{1,2}$ momenta of the incoming parton, as well as on the renormalization ($\mu$) and rapidity ($\zeta$) scales, here set as $\mu^2=\zeta_1=\zeta_2=Q^2$. 
Most importantly, they contain an intrinsic non-perturbative core that becomes increasingly relevant as $|\kperp|$ decreases.
This is reflected in the DY spectrum at small $|\qT|$, which requires the introduction of non‑perturbative physics, in contrast with the regime $|\qT| \sim Q$ where the transverse-momentum dependence is generated perturbatively~\cite{Collins:1984kg,Bozzi:2005wk,Collins:2011zzd,Aybat:2011zv}.
Such scenario poses the difficult challenge of separating genuine non-perturbative contributions from those parts that can still be treated using perturbative techniques. 
This is a central objective in TMD phenomenology. 
In fact, most efforts are dedicated to disentangle the non-perturbative core of TMD parton densities, extract it from data, and interpret it in terms of physically meaningful functions encoding the intrinsic three-dimensional motion of partons in hadrons. 
Ideally, the methodology should remain independent of arbitrary assumptions, with different parameterizations eventually converging to the same physical description. 
However, modeling non-perturbative physics inevitably builds on choices, assumptions and prescriptions whose effects are difficult to control, leading to extracted quantities that represent a combination of different intertwined effects. 
This can potentially obscure the interpretation of the outcome of the phenomenological analysis, although still providing an excellent agreement with experimental data.   
Reducing the phenomenological bias is therefore an important concern, particularly in view of the tensions currently reported among TMD extractions by different groups (see Ref.~\cite{Moos:2025sal}).

The foundations of TMD phenomenology traces back to the Collins–Soper–Sterman (CSS) formalism~\cite{Collins:1981va,Collins:1984kg}, which will be reviewed in Section~\ref{ss:standard}. 
This approach has been extensively developed and refined, reaching very high perturbative accuracy, up to next-to-next-to-next-to-leading log N$^3$LL (and approximate N$^4$LL for Drell–Yan processes) in the most recent studies~\cite{Moos:2025sal,Bacchetta:2025ara,Bacchetta:2024qre,Moos:2023yfa,Bacchetta:2022awv,Rossi:2025pwh,Barry:2025glq}. 
In these studies, the parameterization of the non-perturbative core of TMDs has reached high level of complexity, and only in some cases it is possible to extract physically interpretable quantities like the average partonic transverse momentum, which often depend on procedures based on specific assumptions (see, \textit{e.g.}, Sec.~IV~B.4 of Ref.~\cite{Bacchetta:2024qre}).
More recently, in order to reduce the potential bias associated with model choices, neural networks have been successfully tested to parametrize the non-perturbative components of TMDs~\cite{Bacchetta:2025ara,Avkhadiev:2025wps}. 
This strategy is shown to improve the description of the data and provide more reliable error bands, increasing the confidence that the underlying physics is contained within the extracted uncertainties. 
The limitation of this approach is the loss of direct interpretability of the results in terms of QCD mechanisms, which remains one of the strengths of more traditional parameterizations. 
Alternative approaches~\cite{Gonzalez-Hernandez:2023iso,Aslan:2024nqg,Simonelli:2025kga}, which aim to maximize the interpretability of the results, have been proposed only recently and are therefore still at an early stage of phenomenological development. 
At present, the original CSS approach remains the most widely used and extensively adopted framework for TMD extractions.

The purpose of this work is to assess the impact of potential biases inherent in the phenomenological implementation of the CSS approach. We provide explicit examples by performing fits to a selection of Drell–Yan experimental data and discuss the practical advantages and intrinsic limitations of this standard approach. Specifically, we present a detailed analysis of the effects of the phenomenological assumptions introduced by the so-called ``$\bstar$ prescription'' to avoid hitting the Landau pole. 
We emphasize that increasing awareness of the potential limitations of the standard strategy allows one to exploit its intrinsic high flexibility for a reliable reconstruction of the general TMD behavior.
Finally, we show that performing a global fit combining low- and high-energy data helps reduce phenomenological biases, significantly improving the stability of the extraction.

\section{Resummation and Non-Perturbative modeling}
\label{s:formalism}

The typical strategy to address the factorization theorem of Eq.~\eqref{e:DYZ_xsec} is to exploit perturbative QCD wherever applicable and to supplement it with a non-perturbative component when it starts to break down. Starting from expansions in the strong coupling, one immediately realizes that large logarithms $L = \log(Q/|\qT|)$ arise order by order as the transverse momentum decreases, spoiling convergence of fixed-order theory and signaling the presence of non-perturbative dynamics. 
This issue is directly visible in the TMD distributions. Indeed, their expression in powers of the strong coupling is an operator-product-expansion (OPE) in terms of usual PDFs~\cite{Collins:2011zzd,Echevarria:2011epo,Echevarria:2016scs,Scimemi:2019gge,Aslan:2024nqg}:
\begin{align}
\label{eq:tmd_NLO_kTspace}
    &F_{q/h}(x,|\kperp|,\mu,\zeta/\mu^2) = 
    \frac{a_S(\mu)}{2\pi |\kperp|^2}
    \int_x^1 \frac{d \hat{x}}{\hat{x}}
\notag \\
&\quad
    4 C_F \Big[
     \frac{(1 + \hat{x}^2)}{(1-\hat{x})_+} +\log{\big({\zeta}/{|\kperp|^2}\big)}\delta(1-\hat{x})
   \Big]\, f_{q/h}(x/\hat{x};\mu)
+
   2\Big[\hat{x}^2 + (1-\hat{x})^2 \Big]\,f_{g/h}(x/\hat{x};\mu)
    +
   \mathcal{O}(a_S^2)  \, ,
\end{align}
which is valid only within the narrow kinematic region $|\kperp| \approx \mu$, or, at the level of the cross section, $|\qT| \approx Q$. Note that distributional terms that would arise from a naive perturbative extension into the low-$|\kperp|$ region, such as contributions involving $\delta(|\kperp|)$ are beyond the validity domain of this OPE.
Perturbative predictions can nevertheless be extended towards lower values of transverse momentum by resumming the large transverse-momentum logarithms.
This procedure captures all-order towers of contributions, such as the leading-logarithmic (LL) terms, proportional to $a_s^n L^{2n-1}/|\qT|^2$, the next-to-leading-logarithmic (NLL) terms, proportional to $a_s^n L^{2n-2}/|\qT|^2$, and so on.
Carrying out this resummation directly in transverse-momentum space, however, proves extremely challenging. 
Severe obstacles are known to arise already at NLL accuracy~\cite{Frixione:1998dw}, whose solution requires a careful analytic analysis~\cite{Simonelli:2025kga}. 
The resummation procedure becomes considerably simpler in the impact-parameter space $\bT$, Fourier conjugate to $\qT$, where the convolution in Eq.~\eqref{e:DYZ_xsec} simplifies to ordinary products, yielding a particularly transparent form of the factorization theorem:
\begin{align}
    \label{eq:tmd_fact_b}
    &\frac{d\sigma^{\text{DY}}}{d |\qT|\, dy\, dQ} = \frac{16 \pi^2 \alpha^2 |\qT|}{9 Q^3}\, x_1\, x_2 \, {\cal H}^{\text{DY}}\big(a_S(\mu);{Q}/{\mu}\big) \, \sum_q c_q(Q^2)
    \int \frac{d^2\bT}{(2\pi)^2} e^{-i \qT \cdot \bT} 
    \notag \\
    &\qquad\times
    \fourier{F}_{q/h_1}\big(x_1, \bT^2; \mu, \zeta_1\big)
    \fourier{F}_{\overline{q}/h_2}\big(x_2, \bT^2; \mu, \zeta_2\big)
    + (1  \leftrightarrow 2)  \, ,
\end{align}
in terms of the Fourier transformed TMD distributions $\fourier{F}_{q/h_i}$.
In this auxiliary space, the large logarithms plague the perturbative expansion at large distances, appearing in the distributions as $L_b = \log(\mu |\bT|/c_1)$, with $c_1 = 2 e^{-\gamma_E} \simeq 1.123$ and $\gamma_E$ the Euler–Mascheroni constant.
However, here they can be easily resummed, stretching the applicability  of perturbative QCD from the narrow region $|\bT| \approx c_1/\mu$ up to distances of $|\bT| \approx 1\,\text{GeV}^{-1}$. In fact, all logarithmic contributions exponentiate into a well-defined structure (see, \textit{e.g.}, Ref~\cite{Bozzi:2010xn}):
\begin{align}
    &\fourier{F}_{q/h}(x,\bT^2;\mu) = 
    f_{q/h}(x,\mu_b) \,
    \text{exp}\Big\{
    L_b g_1(\lambda_b) +  g_2(\lambda_b) + \frac{1}{L_b} g_3(\lambda_b) + \dots
    \Big\} \, ,
    \label{eq:TMD_solEVO_resumm}
\end{align}
where we have implicitly set $\zeta=\mu^2$ and defined $\lambda_b = 2 \beta_0 a_S(\mu) L_b$, with $\beta_0$ being the lowest order coefficient of the QCD beta function. We have also introduced the scale $\mu_b=c_1/|\bT|$ at which the logarithms $L_b$ vanish. The functions $g_i$ encode towers of logarithms to all orders, with $g_1$ reproducing the LL terms $a_S^n L_b^{2n}$, $g_2$ the NLL terms $a_S^n L_b^{2n-1}$, and so on. 
This result traces back to the particularly simple form of the evolution equations~\cite{Collins:2011zzd,Aybat:2011zv,Echevarria:2011epo,Boussarie:2023izj}:
\begin{subequations}
\label{eq:TMD_evo}
\begin{align}
\label{eq:TMDPDF_RGevo}
    &\frac{\partial}{\partial \log{\mu}} \log{\fourier{F}_{q/h}(x,\bT^2;\mu,\zeta)}
    =
   \gamma_f \big(a_S(\mu)\big) - \frac{1}{2}\gamma_K\big(a_S(\mu)\big) \, \log{\big({\zeta/\mu^2}\big)} \, ,
    \\
\label{eq:TMDPDF_CSevo}
    &\frac{\partial}{\partial \log{\sqrt{\zeta}}} \log{\fourier{F}_{q/h}(x,\bT^2;\mu,\zeta)}
    =
    K\big(a_S(\mu); |\bT| \mu/c_1\big) \, ,
\end{align}
\end{subequations}
where $\gamma_f$ denotes the typical TMD PDF anomalous dimension, while $K$ is the Collins–Soper (CS) kernel, which governs the dependence on the rapidity scale and evolves with the cusp anomalous dimension $\gamma_K$. The general solution of TMD evolution equations is 
\begin{align}
    &\fourier{F}_{q/h}(x,\bT^2;\mu,\zeta) = \fourier{F}_{q/h}(x,\bT^2;\mu_b,\mu_b^2)
    \notag \\
    &\quad\times
    \text{exp}\Big\{
    \log{\Big(\frac{\mu}{\mu_b}\Big)} K\big(a_S(\mu_b); 1\big)
    +
    \int_{\mu_b}^\mu \frac{d \mu'}{\mu'}
    \Big[
    \gamma_f\big(a_S(\mu')\big)
    -\gamma_K\big(a_S(\mu')\big)\,
    \log{\frac{\mu}{\mu'}}
    \Big]
    \Big\}
   \notag \\
   &\quad\times
    \text{exp}\Big\{
    \log{\Big(\frac{\sqrt{\zeta}}{\mu}\Big)} K\big(a_S(\mu); |\bT| \mu/c_1\big)
    \Big\} \, .
    \label{eq:TMD_solEVO}
\end{align}
This leads to Eq.\eqref{eq:TMD_solEVO_resumm} upon exploiting the OPE in Fourier space and analytically solving the integrals over the anomalous dimensions; both approximations are valid as long as $\mu_b$ remains a perturbative scale, namely as $|\bT|$ is small. 
Below this threshold, the emergence of non-perturbative effects is unavoidable, as signaled by the branch cut at $\lambda_b=1$ in the functions $g_i$ of Eq.\eqref{eq:TMD_solEVO_resumm}. This is the manifestation of the Landau pole, indicating the breakdown of perturbative tools.
Its treatment is connected to the description of the non-perturbative regime. Moreover, in impact-parameter space, the introduction of an appropriate model is unavoidable, since the inverse Fourier transform requires the distributions across the entire $|\bT|$ range, including very large distances.

An alternative strategy has been proposed in Ref.~\cite{Aslan:2024nqg}, where the non-perturbative core is treated as the central ingredient and fixed-order perturbative expansions are matched onto it. However, this approach neglects the region at intermediate transverse momentum, where fixed-order calculations are insufficient to reliably describe TMD distributions, while perturbative tools remain applicable through resummation.
For this reason, common strategies set up a prescription to avoid the Landau cut while keeping into account the resummation, and then encode all non-perturbative effects in unknown multiplicative factors that must be determined by fitting experimental data (see, \textit{e.g.}, Eq.~(2.36) of Ref.~\cite{Bacchetta:2019sam}).  
This approach allows for significant flexibility in the implementation of non-perturbative effects. This flexibility is advantageous for achieving accurate data descriptions, but it can also complicate the separation of genuine non-perturbative physics from artifacts associated with the chosen prescription to regulate the Landau region. As a result, the physical interpretation of the extracted functions may become less transparent, even when the overall agreement with data is excellent. 

\subsection{The standard approach}
\label{ss:standard}

The Landau cut that appears when applying perturbative methods to Eq.~\eqref{eq:TMD_solEVO} occurs at $|\bcut| = c_1/\mu \, \exp\big(1/{2 \beta_0 a_S(\mu)}\big)$, which typically corresponds to much larger distances than the characteristic scale where genuine non-perturbative effects are expected to emerge, around 1~GeV$^{-1}$.
This scale is a consequence of the extension of perturbation theory beyond its natural domain, and it requires the introduction of a suitable prescription in order to be consistently treated. 
Among the various prescriptions proposed in the literature~\cite{Catani:1996yz,Laenen:2000de,Kulesza:2002rh,Bonvini:2008ei}, the $\bstar$ prescription is by far the most widely adopted in TMD phenomenology, tracing back to seminal works in the 80's by Collins, Soper, and Sterman~\cite{Collins:1981va,Collins:1984kg}. It consists in mapping the branch cut to infinity through a suitable redefinition of $\bT$, via a function $\bstar(|\bT|)$ that saturates to a limiting value $\bmax$ as $|\bT| \to \infty$, while coinciding with $|\bT|$ in the opposite limit.
Apart from these asymptotic properties, the functional form of $\bstar$ is arbitrary, as well as the specific value of $\bmax$, which is only required to remain below $|\bcut|$.

Therefore, such prescription modifies the TMD distributions at large distances, allowing for a safe evaluation of the inverse Fourier transform.
This approach, however, goes beyond a purely technical prescription and it effectively acts as an infrared model governing the behavior of TMD distributions beyond the perturbative regime.
Ref.~\cite{Simonelli:2025kga} showed that this effectively parametrizes the deep-infrared behavior of both the strong coupling and the collinear PDFs.
Consequently, the logarithms in Eq.~\eqref{eq:TMD_solEVO_resumm} are prevented from growing indefinitely and freeze at distances of order $\bmax$ (or, equivalently, at scales of order $c_1/\bmax$).

Phenomenological studies can proceed along two complementary directions. One option is to embrace the interpretation of the $\bstar$ prescription as part of the non-perturbative behavior of TMD distributions, refining it if necessary to construct more sophisticated infrared models, as recently explored in Ref.~\cite{Simonelli:2025kga}. Alternatively, one can reject the idea that the TMDs depend on the $\bstar$ prescription and attempt to construct distributions that are, by definition, independent of the specific choice, assigning the entire role of parametrizing the infrared behavior to suitable non-perturbative factors.
This second option corresponds to the approach originally proposed by Collins, Soper, and Sterman in Refs.~\cite{Collins:1981va,Collins:1984kg} and is hereafter referred to as the \emph{standard approach}. 
It is based on the principle that $\bmax$ is an auxiliary, unphysical parameter introduced solely by the prescription, and, therefore, any physical prediction should be independent of it. To achieve this, the non-perturbative components need to be defined so that their dependence on $\bmax$ exactly compensates the dependence introduced in the perturbative regime.
This is implemented via two scale-invariant non-perturbative functions, $g_K$ and $g_{q/h}$, which can be regarded as the long-distance counterparts of the perturbative $g_i$ functions appearing in the resummed expression of Eq.~\eqref{eq:TMD_solEVO_resumm}. 
The function $g_K$ encodes the non-perturbative behavior of the CS kernel:
\begin{align}
\label{eq:gK_def}
&g_K(|\bT|;\bmax) = K\big(a_S(\mu), \bstar(|\bT|) \mu/c_1 \big) - K\big(a_S(\mu), |\bT| \mu/c_1 \big) \, ,
\end{align}
while the function $g_{q/h}$ encodes the long-distance behavior of the specific TMD and it can be regarded as its ``intrinsic'' part:
\begin{align}
\label{eq:gF_def}
&g_{q/h}(x,|\bT|;\bmax) 
+ \log{\Big({\sqrt{\zeta}}/{Q_0}\Big)} g_K(|\bT|;\bmax) 
= \log{\Bigg(\frac{\fourier{F}_{q/h}(x,\bstar(|\bT|);\mu,\zeta/\mu^2)}{\fourier{F}_{q/h}(x,|\bT|;\mu,\zeta/\mu^2)}\Bigg)} \, ,
\end{align}
where $Q_0$ is an arbitrary input rapidity scale. 
These definitions leads to the well-known expression for the TMD distribution:
\begin{align}
    &\fourier{F}_{q/h}(x,|\bT|;\mu,\zeta/\mu^2) = \fourier{F}_{q/h}(x,\bstar(|\bT|);\mustar,1)
    \notag \\
    &\quad\times
    \text{exp}\Big\{
    \log{\Big(\frac{\mu}{\mustar}\Big)} K\big(a_S(\mustar); 1\big)
    +
    \int_{\mustar}^\mu \frac{d \mu'}{\mu'}
    \Big[
    \gamma_f\big(a_S(\mu')\big)
    -\gamma_K\big(a_S(\mu')\big)\,
    \log{\Big({\mu}/{\mu'}\Big)}
    \Big]
    \Big\}
   \notag \\
   &\quad\times
    \text{exp}\Big\{
   -g_{q/h}(x,|\bT|;\bmax) 
- \log{\Big({\sqrt{\zeta}}/{Q_0}\Big)} g_K(|\bT|;\bmax) 
    \Big\}
    \label{eq:TMD_solEVO_std}
\end{align}
This is simply an explicit reformulation of the general solution of the TMD evolution equations and is therefore manifestly independent of $\bmax$.
This rewriting formally assigns all non-perturbative effects to the functions $g_K$ and $g_{q/h}$, whose assumed functional forms must guarantee $\bmax$ independence.

Unfortunately, this requirement strongly limits their flexibility and prevents their use in phenomenological applications. In fact, $g_K$ and $g_{q/h}$ are often implemented through relatively simple models (for instance quadratic forms), rarely including the explicit $\bmax$ dependence required to satisfy the condition discussed above\footnote{In Refs.~\cite{Bertone:2019nxa,Scimemi:2019cmh,Bury:2022czx,Moos:2023yfa,Moos:2025sal,Barry:2025glq,Camarda:2025lbt}, the parameterization of the $g_K$ function contains a dependence on $\bstar$ that, however, does not cancel the same dependence in the perturbative counterpart.}. As a result, the TMDs do acquire an implicit dependence on $\bmax$, which can have a non-negligible impact. In this context, treating $\bmax$ as a fixed, non-fitted parameter becomes less justified, as it influences the long-distance behavior of the distributions.
Nonetheless, it is common to fix $\bmax$ and treat the PDFs as external inputs. Phenomenological efforts then focus on modeling $g_K$ and $g_{q/h}$, assuming that the effect of the $\bstar$ prescription on the non-perturbative sector is small. A natural choice for $\bmax$ is $c_1$, since this ensures that the scale $\mustar$ always remains within the perturbative domain, with $a_S$ evaluated above $1~\text{GeV}$. 
In this way, one avoids dealing with PDFs below this threshold, where they are not tabulated.
For these reasons, the impact of $\bmax$ and of the $\bstar$ prescription on phenomenology remains to be critically tested, with the aim of assessing how these choices influence the extraction and interpretation of TMDs.

The parametrization of the transition between the perturbative and non‑perturbative regimes is only one possible source of tension between existing TMD extractions. Another potentially impacting methodological choice concerns the implementation of the initial rapidity scale in Eq.~\eqref{eq:TMDPDF_CSevo}. In this work, following the MAP collaboration, we use $\zeta = Q^2$. 
An alternative approach is the $\zeta$-prescription, based on the concept of “optimal” TMDs~\cite{Scimemi:2017etj,Scimemi:2018xaf}.
Although this choice is expected to have a negligible effect on the cross section calculation, it may still influence the behavior of TMD distributions at low transverse momentum. We postpone the analysis of this issue to future work and focus here solely on the effects of the $\bstar$ implementation.
Finally, the choice of collinear PDFs can have a non-negligible effect on TMD phenomenology. In this work, we focus on kinematic regions where PDFs are well constrained and consistent among different sets, ensuring that this source of uncertainty has minimal impact on our results.

\section{A phenomenological stress-test}

Low energy Drell-Yan data sets \textsc{E288}~\cite{Ito:1980ev} and \textsc{E605}~\cite{Moreno:1990sf} represent a convenient testing ground for studying the impact of the phenomenological choices at the foundation of the standard approach. The low-$|\qT|$ behavior of these data has been successfully described since many years ago
and it is also currently very well described by modern TMD approaches~\cite{Sun:2014dqm,Bacchetta:2017gcc,Bertone:2019nxa,Bacchetta:2019sam,Bacchetta:2022awv,Aslan:2024nqg,Moos:2023yfa,Bacchetta:2024qre,Bacchetta:2025ara,Moos:2025sal,Barry:2025glq,Fernando:2025xzv,Camarda:2025lbt}.
Our strategy is to adopt a simple parametrization for the $g$-functions and keep it fixed while varying the specific choices involved in the $\bstar$ prescription. This allows us to estimate the impact of these variations in a realistic scenario and, ultimately, to assess to what extent the standard approach enables us to probe the internal structure of hadrons.  

\subsection{Analysis setup}
\label{ss:Pheno_setup}

We illustrate in this section the phenomenological choices made for all the fits discussed in the following. The logarithmic accuracy of our analysis is set to next-to-next-to-leading order (NNLL).
Although this is not the current highest accuracy, we believe that it is enough for the goal of this work.
Indeed, it has been shown in the literature that low-energy DY data can be described at this perturbative accuracy (see, \textit{e.g.}, Ref.~\cite{Bacchetta:2022awv}). This allows for an extraction of TMD PDFs that can be consistently matched to NLO perturbative calculations (for more details, see Tab.~I of Ref.~\cite{Bacchetta:2022awv}).

Another important choice that is included in the phenomenological setup is the set of collinear parton distributions appearing in the operator product expansion at large transverse momentum (see Eq.\eqref{eq:tmd_NLO_kTspace}).
We choose the PDFs set \textsc{MMHT2014}~\cite{Harland-Lang:2014zoa}. This choice is comparable with the ones made in previous works~\cite{Bacchetta:2019sam,Bacchetta:2022awv,Aslan:2024nqg}.

As figure of merit for the assessment of the data/theory agreement we adopt the $\chi^2$ function as defined in recent TMD analyses (see, \textit{e.g.}, Eq.~(55) of Ref.~\cite{Bacchetta:2022awv}), including the penalty term due to correlated uncertainties. It is built by nuisance parameters obtained from the minimization of the full $\chi^2$ on data (see Refs.~\cite{Ball:2008by,Ball:2012wy,Bertone:2019nxa,Bacchetta:2019sam} for more details).
The propagation of the errors is performed through the so-called bootstrap method, which consists in fitting an ensemble of Monte Carlo (MC) replicas of the experimental data (see Ref.~\cite{Bacchetta:2017gcc} for more details). We choose the number of replicas by requiring that average and standard deviation of the ensemble accurately reproduces the original data points. In this case, we find that 300 replicas are sufficient.

The phenomenological results presented in this work are obtained with the public version of the \textsc{NangaParbat}~\cite{Bacchetta:2019sam,Bacchetta:2022awv} fitting framework.

\subsubsection{Included dataset}
\label{ss:data}

Among the full DY dataset included in state-of-the-art TMD extractions, we consider measurements from the E605~\cite{Moreno:1990sf} and E288~\cite{Ito:1980ev} experiments at Fermilab. They cover a relatively narrow kinematic range in the longitudinal momentum fraction $x$, which justifies a very simple choice for the TMD non-perturbative model as in Eq.~\eqref{e:fNP}. A detailed description of the features of these data sets can be found, \textit{e.g.}, in Refs.~\cite{Bacchetta:2019sam,Aslan:2024nqg}. 
Since our analysis is based on the applicability of TMD factorization, which is valid only in the region $|\qT| \ll Q$, we impose the following cut on the transverse momentum of the intermediate photon:
\begin{equation}
\label{e:DYcut}
|\qT| < 0.2\, Q \, .
\end{equation}
Moreover, we also excluded the bins at $9.00 < Q < 11.70 \text{~GeV}$ to avoid the kinematical region of invariant masses around the $\Upsilon$ resonance. In total, we fit 180 data points (130 for \textsc{E288} experiment and 50 for \textsc{E605}).

Each of the considered data sets is affected by systematic and statistical uncertainties. We treat the former as fully correlated, the latter as uncorrelated. Additionally, we introduce the theoretical systematic uncertainties arising from the collinear PDFs set in the same approach as previous works by the MAP collaboration~\cite{Bacchetta:2022awv,Bacchetta:2024qre,Bacchetta:2024yzl,Bacchetta:2025ara}.

\subsubsection{Choice of non-perturbative model}
\label{ss:g-fun}

The simple parameterization of the $g$-functions in Eq.~\eqref{eq:TMD_solEVO_std} consists in the following non-perturbative models. 
The long-distance behavior of the CS-kernel is parametrized as:
\begin{equation}
\label{e:gK}
g_K(\bT^2) = g_2\, \frac{\bT^2}{2} \, ,
\end{equation}
where $g_2$ is a free parameter. The reference non-perturbative scale $Q_0$ associated to the $g_K$ term in Eq.~\eqref{eq:gF_def} is set to $Q_0^2=3.2$~GeV$^2$. Differences in the choice of $Q_0^2$ are fitted out in the value of the parameter $g_2$.

The ``intrinsic'' non-perturbative corrections encoded into $g_{j/h}$ generally depends on $x$ and $\bT^2$, and possibly the partonic flavor and the hadron species. We use the following simple gaussian model, that only depends on $b_T$: 
\begin{equation}
\label{e:fNP}
f_{1NP}(\bT^2) = e^{-g_1\frac{\bT^2}{2}} \equiv e^{-g_{j/h}(b_T)},
\end{equation}
with $g_1$ a positive free parameter. 
In total, the models involve 2 free parameters, one for the CS kernel and one for the intrinsic part of the TMD. 

\subsubsection{$\bstar$-prescriptions under consideration}
\label{ss:bstar_choices}

There are many ways to choose the $\bstar$-prescription functional form to be included in the standard CSS approach. In this work we focus on the following two most common choices for the function $\bstar(\bT^2)$.
First, we consider the functional form that was originally proposed in Ref.~\cite{Collins:1984kg}: 
\begin{equation}
\label{e:bstar_collins}
\bstar^{\rm CSS}(\bT^2) = \sqrt{\frac{\bT^2}{1 + \frac{\bT^2}{\bmax^2}}} \, ,
\end{equation}
In recent years, another expression has been proposed (see Ref.~\cite{Bacchetta:2015ora})
\begin{equation}
\label{e:bstar_echevarria}
\bstar^{\rm exp}(\bT^2) = \bmax [ 1 - \exp{ (- \bT^4 / \bmax^4)} ]^\frac{1}{4} \, ,
\end{equation}
which leads to a sharper transition between the perturbative and the non-perturbative regions.
\begin{figure}[h!]
\centering
\includegraphics[width=0.55\textwidth]{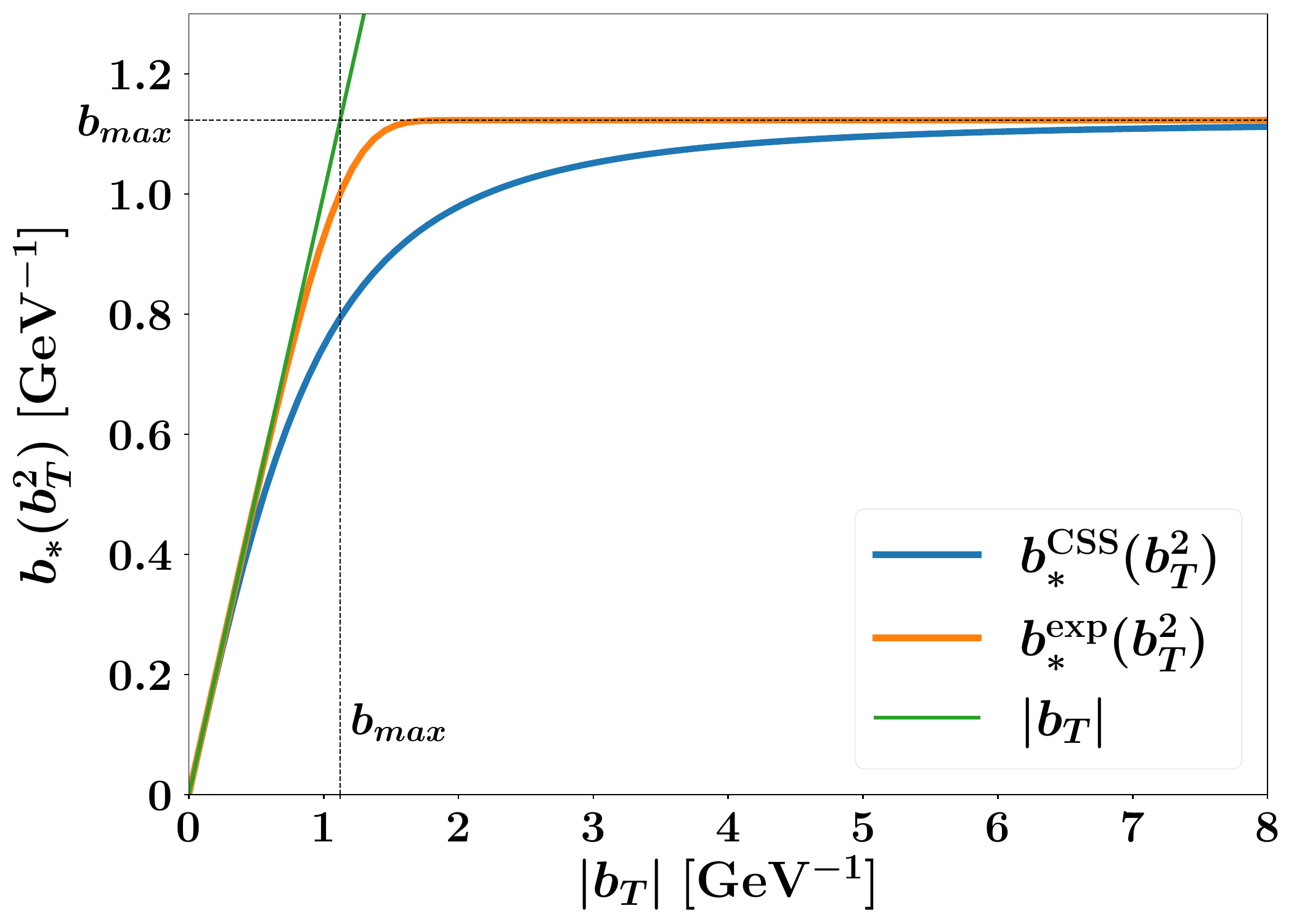}
\caption{Comparison between the definition of the $b_*(\bT^2)$ functions in Eqs.~\eqref{e:bstar_collins}-\eqref{e:bstar_echevarria}, represented in blue and orange, respectively. We fixed $\bmax =  2e^{-\gamma_E}$ (with $\gamma_E$ the Euler constant).}
\label{f:bstar}
\end{figure}
In Fig.~\ref{f:bstar}, the comparison between the two functions in Eqs.~\eqref{e:bstar_collins}-\eqref{e:bstar_echevarria} is displayed. While the small-$|\bT|$ limit of the two functions is exactly the same, the function $\bstar^{\rm CSS}(\bT^2)$ approaches the non-perturbative region in a milder way than $\bstar^{\rm exp}(\bT^2)$, and it deviates from the behavior of $|\bT|$ earlier.

We also test different choices for the value of $\bmax$ and how it correlates with the specific functional form. First, we adopt the square-root expression of Eq.~\eqref{e:bstar_collins}, together with the standard value $\bmax = c_1$, which roughly sets the boundary of non-perturbative effects below $q_T \approx 1$ GeV. Then, we consider the choice, labeled as ``mod'' in our fits, $\bmax = c_1/2$, that effectively raises the non-perturbative threshold to about 2 GeV. 
Next, we repeat our analysis adopting the functional form of Eq.~\eqref{e:bstar_echevarria}. This set-up leads to a total of four different combinations that we can test, summarized in Tab.~\ref{tab:fits}.
\begin{table}[h!]
\centering
\label{tab:fits}
\setlength{\tabcolsep}{10pt} 
\begin{tabular}{l c r}
\toprule
Fit Name & $\bstar$ functional form & $\bmax$ value \\
\midrule
$\bstar^{\text{\scriptsize CSS}}$      & \multirow{2}{*}{Eq.~\eqref{e:bstar_collins}}     & $c_1$ \\
$\bstar^{\text{\scriptsize CSS-mod}}$  &                                                   & $c_1/2$ \\
\midrule
$\bstar^{\text{\scriptsize exp}}$      & \multirow{2}{*}{Eq.~\eqref{e:bstar_echevarria}}  & $c_1$ \\
$\bstar^{\text{\scriptsize exp-mod}}$  &                                                   & $c_1/2$ \\
\bottomrule
\end{tabular}
\caption{Configurations tested in our fits.}
\end{table}
The results of these four fits can shed light on the theoretical systematic uncertainty due to the choice of the $\bstar$-prescription. 

\subsection{Results of the fits}
\label{ss:TMDs_lowEnergy}

We now discuss the results of the extracted non-perturbative content of the TMD distributions, when varying the $\bstar$ prescription. 
In Tab.~\ref{t:chitable_bstar}, we report the breakdown of the $\chi^2$ values normalized to the number of data points ($N_{\text{dat}}$) for the included data sets. The most complete statistical information about the fit procedure is given by the full ensemble of 300 replicas.
We use $\chi^2$ value of the best fit to the experimental unreplicated data as the most appropriate indicator to quantify the data/theory agreement.
\begin{table}[h!]
\centering
\begin{tabular}{|p{1.5cm}|C{1.5cm}|C{1.5cm}|C{1.5cm}|C{1.5cm}|C{1.5cm}|}
\hline
\multicolumn{2}{|c|}{} & \multicolumn{4}{c|}{\rule{0pt}{2.5ex} $\chi^2 / N_{\text{dat}}$} \\ \hline
Dataset & $\ndat$ & \rule{0pt}{2.5ex} $\bstar^{\rm CSS}$ & $\bstar^{\text{\scriptsize CSS-mod}}$ & $\bstar^{\rm exp}$ & $\bstar^{\text{\scriptsize exp-mod}}$ \\ \hline
\textsc{E288} & 130 & 0.99 & 0.83 & 1.00 & 0.96  \\ \hline
\textsc{E605} & 50 & 1.66 & 1.78 & 1.40 & 1.85  \\ \hline
Total & 180 & 1.18 & 1.09 & 1.11 & 1.21  \\ \hline
\end{tabular}%
\caption{Summary of $\chi^2$ values and total $\chi^2$ normalized for the number of data points $\ndat$ of the various fits performed in this analysis.}
\label{t:chitable_bstar}
\end{table}
We can clearly see that the quality of the fit is stable among the different configuration of the $\bstar$-prescription we tested. In particular, we observe a $\chi^2$ value very close to 1 for the \textsc{E288} experiment, while a slightly larger for \textsc{E605}.
Although a more conservative cut is expected to improve the quality of the fit for this data set, we believe that our choice is sufficient to restrict the analysis into a safe region where TMD factorization holds. We stress that our results are consistent with recent state-of-the-art global analyses~\cite{Bacchetta:2022awv,Bacchetta:2025ara,Moos:2025sal,Barry:2025glq}.

\begin{figure}[!h]
\centering
\includegraphics[width=0.48\textwidth]{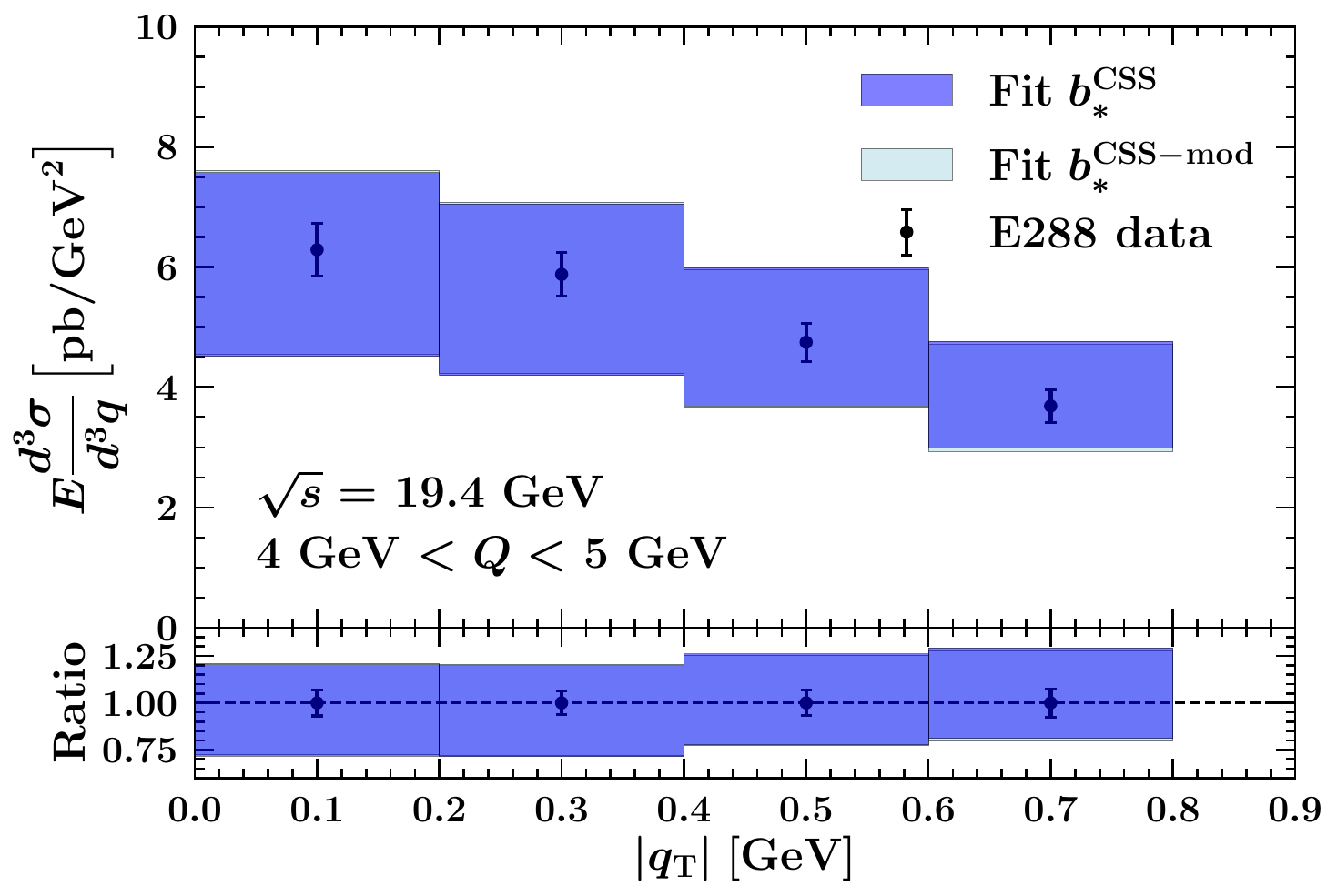}
\includegraphics[width=0.48\textwidth]{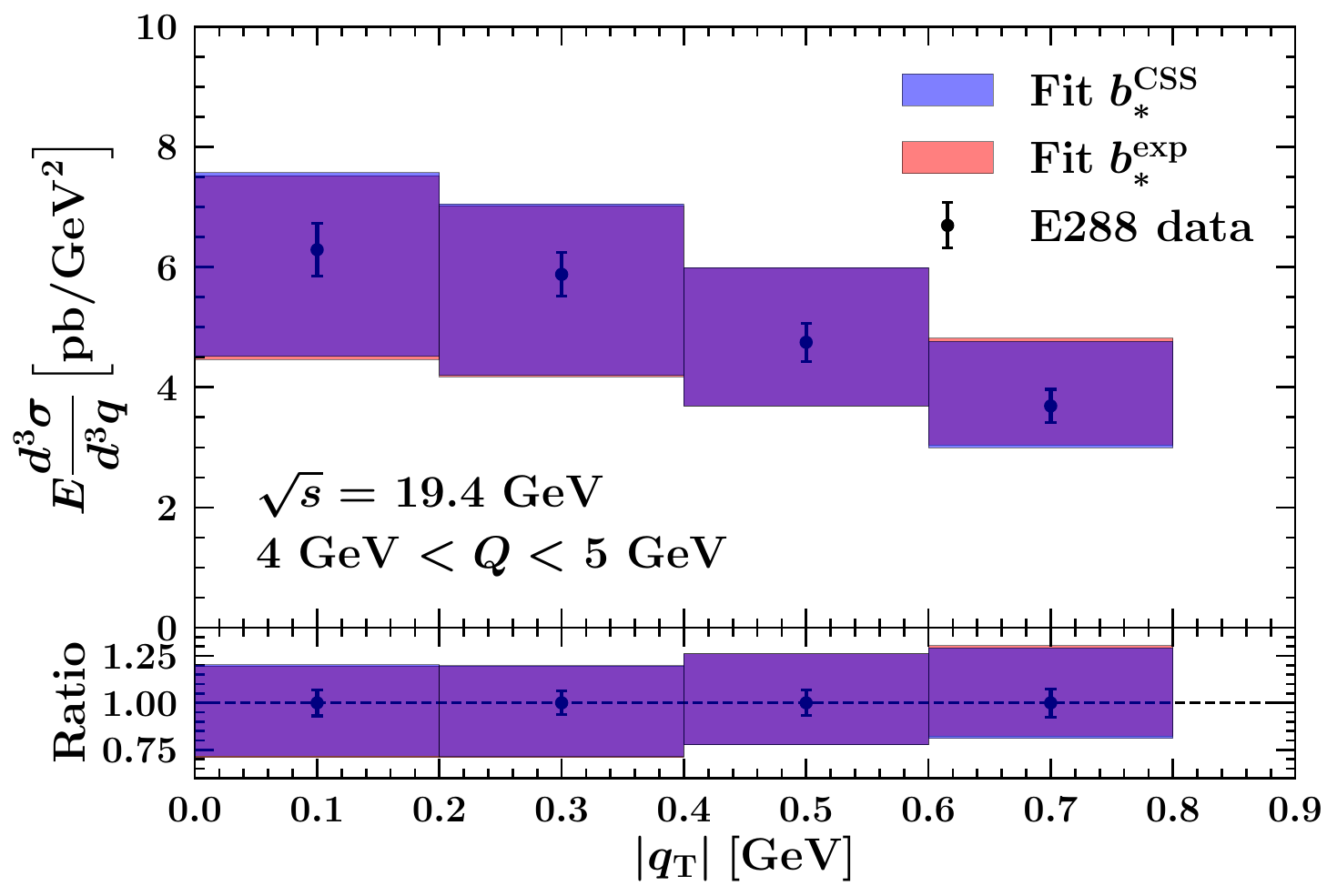}
\includegraphics[width=0.48\textwidth]{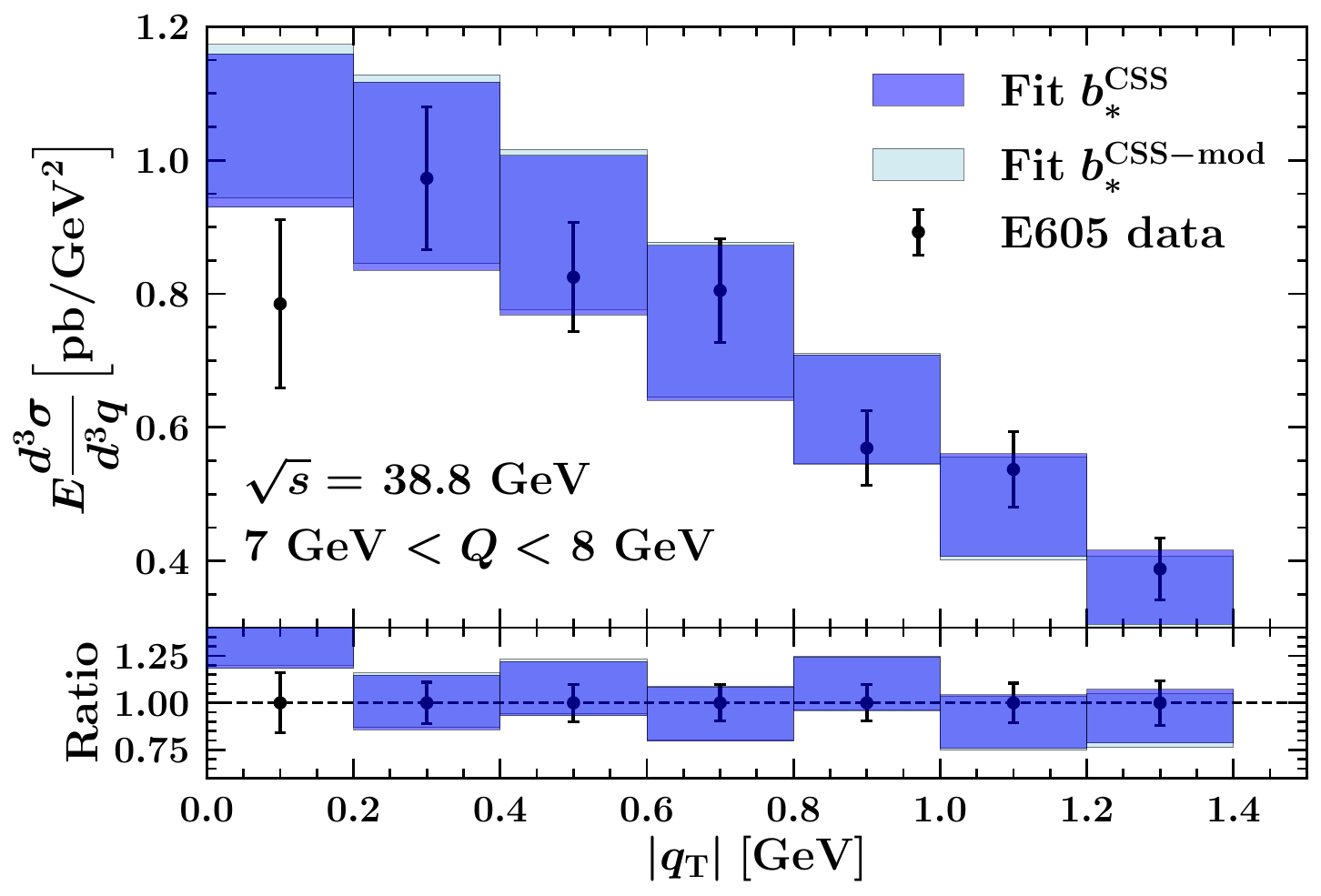}
\includegraphics[width=0.48\textwidth]{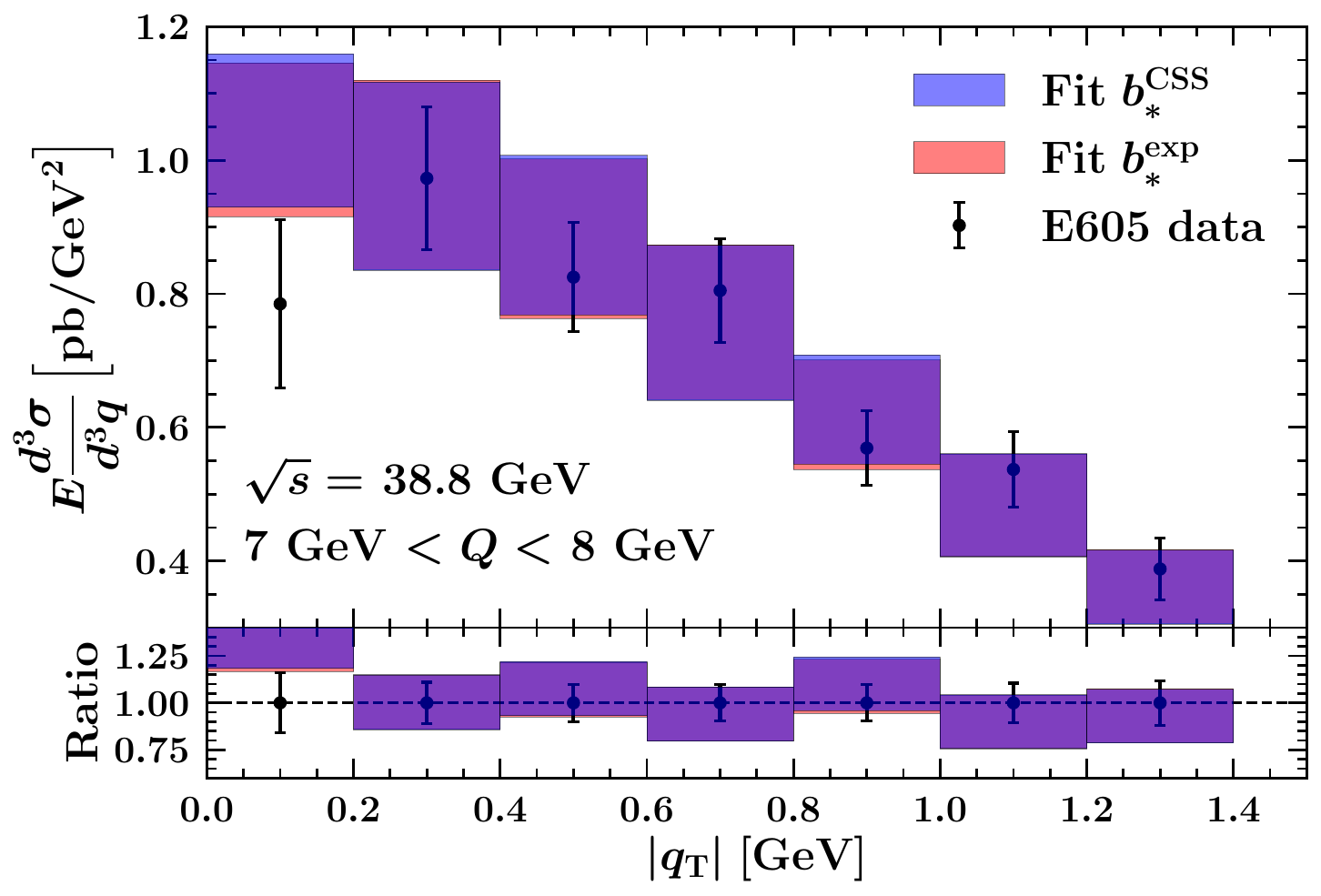}
\includegraphics[width=0.48\textwidth]{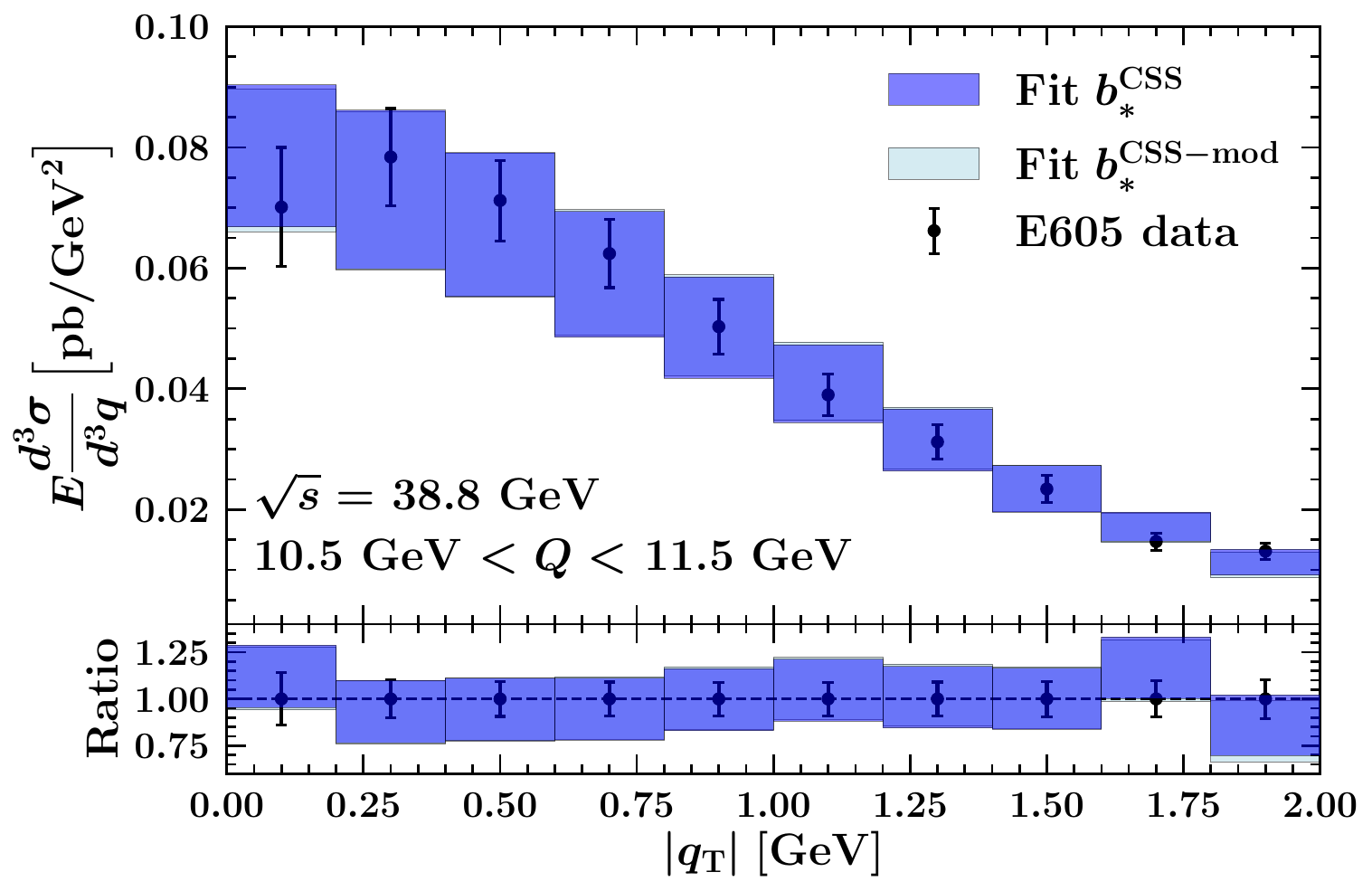}
\includegraphics[width=0.48\textwidth]{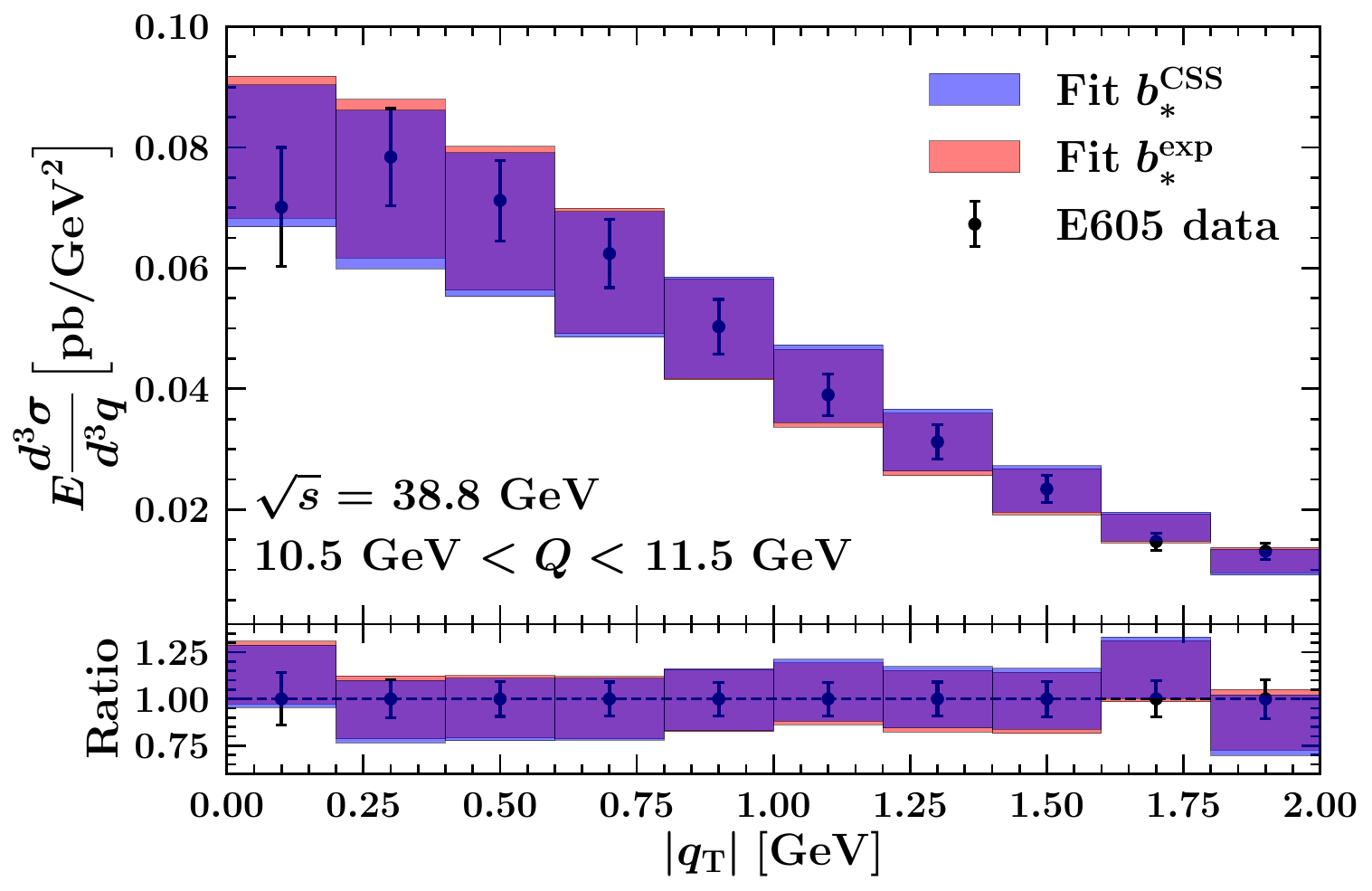}
\caption{
Comparison between data for a selection of $Q$ bins of the \textsc{E288} and \textsc{E605} experiments and theoretical predictions for DY cross section differential in $|\qT|$. Left column, comparison between \textit{Fit $\bstar^{\rm CSS}$} (blue bands) and \textit{Fit $\bstar^{\text{\scriptsize CSS-mod}}$} (light-blue bands); right column: comparison between \textit{Fit $\bstar^{\rm CSS}$} and \textit{Fit $\bstar^{\rm exp}$} (red bands). 
Lower panels: ratio between experimental data and theoretical cross section.
Theoretical uncertainty bands correspond to 68\% C.L., error bars on experimental data display uncorrelated uncertainties only. Single visible color bands signal complete overlap.
}
\label{f:bstar_data-theo}
\end{figure}

In Fig.~\ref{f:bstar_data-theo}, we show the comparison between the result of our fits (colored bands) and experimental data (black points) for a selection of $Q$ bins of the \textsc{E288} and \textsc{E605} data sets. 
In the lower panels, we show the ratio between theory and experimental data. In the left column, we compare the results of \textit{Fit $\bstar^{\rm CSS}$} (blue bands) and \textit{Fit $\bstar^{\text{\scriptsize CSS-mod}}$} (light-blue bands), while in the right column \textit{Fit $\bstar^{\rm CSS}$} and \textit{Fit $\bstar^{\rm exp}$} (red bands). The comparison between \textit{Fit $\bstar^{\rm exp}$} and \textit{Fit $\bstar^{\text{\scriptsize exp-mod}}$} is not reported because the results are qualitatively similar to the ones shown for $\bstar^{\rm CSS}$.
The differential DY cross section is shown as a function of the transverse momentum $|\qT|$ of the virtual vector boson. The uncertainty bands correspond to the 68$\%$ Confidence Level (C.L.).
%
We note very good agreement between data and fitted theory in all the configurations. Particularly, the shape of the experimental data is very well reproduced. 
The broad error bands are a consequence of the large correlated systematic errors of the considered data sets.
Moreover, we observe that the results in the various configurations overlap almost perfectly, showing high stability of the CSS approach in describing the low-energy DY data when the $\bstar$-prescription is modified. Therefore, one might conclude that there is a small dependence on the choice of this ingredient, at least in terms of the agreement between data and theory.
This result reflects the fact that low-energy Drell–Yan data primarily probe the non-perturbative domain of the TMDs, where differences among $\bstar$ prescriptions are effectively absorbed into the fitted parameters.

\bigskip

\begin{figure}[h]
\centering
\includegraphics[width=0.6\textwidth]{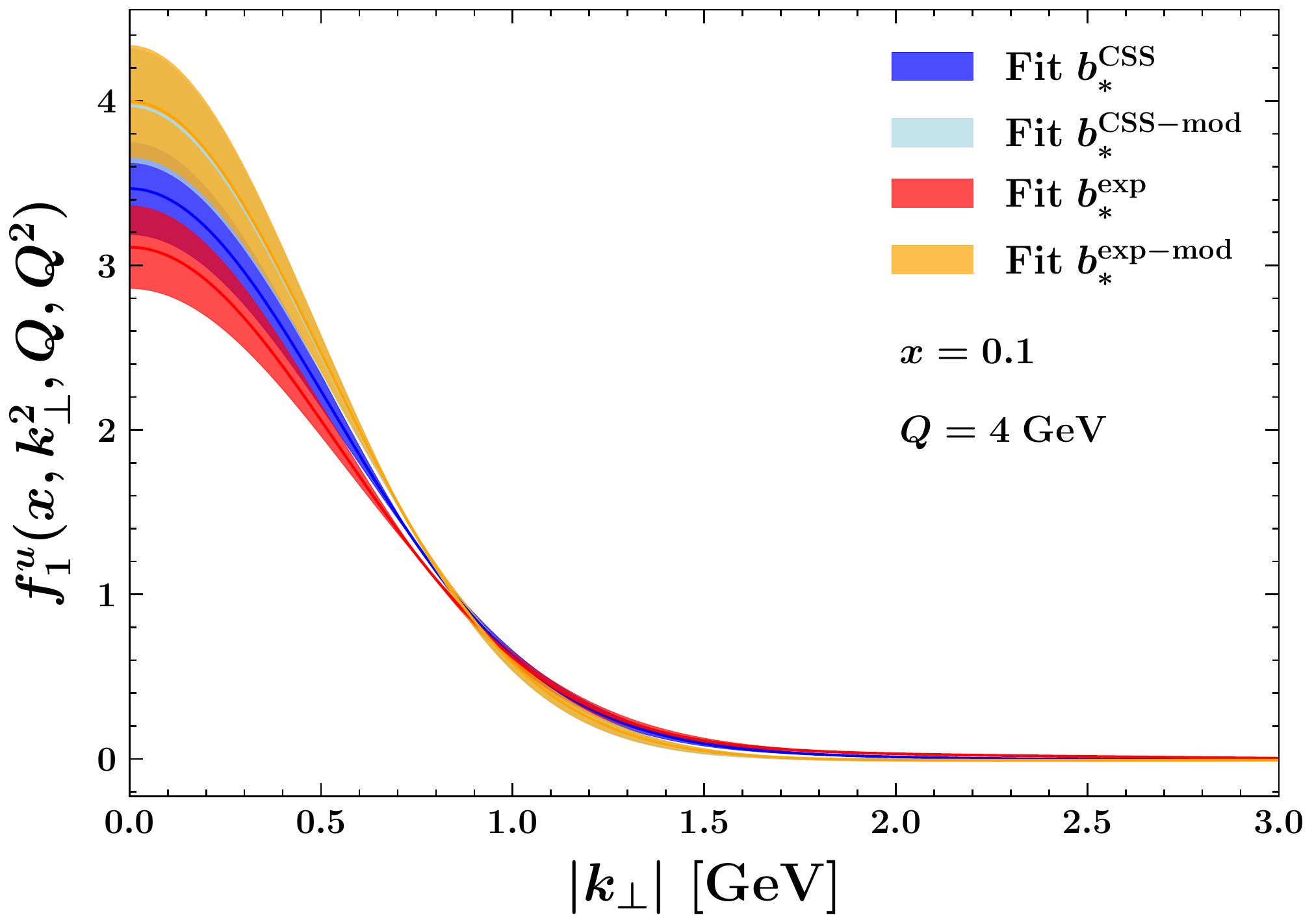}
\caption{The TMD PDF of the up quark in a proton at $\mu = \sqrt{\zeta} = Q = 4$  GeV and $x$ = 0.1 as a function of the partonic transverse momentum $|\kperp|$ as extracted from \textit{Fit $\bstar^{\rm CSS}$} (blue band), \textit{Fit $\bstar^{\rm exp}$} (red band), \textit{Fit $\bstar^{\text{\scriptsize CSS-mod}}$} (light-blue band), and \textit{Fit $\bstar^{\text{\scriptsize exp-mod}}$} (orange band). The uncertainty bands represent the 68\% C.L.}
\label{f:TMDs_bstar}
\end{figure}

This discussion can be extended to the level of the TMD distributions. 
In Fig.~\ref{f:TMDs_bstar}, we show the comparison between the 68$\%$ C.L. error bands of the unpolarized TMD PDF for the up quark in the proton and $x$ = 0.1 at $\mu = \sqrt{\zeta} = Q = 4$ GeV as a function of the quark transverse momentum $|\kperp|$ for the different fits discussed in this section. The value of $x$ and $Q$ are chosen to be consistent with the kinematic region of the dataset included in our analyses.
We observe that the extracted TMDs are compatible within uncertainties across the various fits at low transverse momentum. The different choices for the $\bstar$-prescription introduce only small effects on the non-perturbative part of the TMD distributions obtained in our analysis. This is due to the fact that the included data constrain primarily the small-$|\kperp|$
region of the TMD operator.
In fact, the non-perturbative parameters are well determined and generally compatible across the fit configurations. Overall, the extracted TMDs show a high degree of stability in the region of non-perturbative $|\kperp|$, demonstrating that our results are robust with respect to the specific choice of $\bstar$.

On the other hand, while it is reasonable to expect only mild discrepancies among the TMDs extracted using the four different $\bstar$ prescriptions in the fixed-order region at large $|\kperp| \approx Q$, it cannot be assumed that they behave similarly in the intermediate region ($\text{1-2 GeV} \ll |\kperp| \ll Q$).
In the following section, we show that their behavior in this region is indeed noticeably different, leading to significant variations in the predicted cross sections for high-energy Drell–Yan processes. 

\subsection{TMDs in the intermediate region}
\label{ss:TMDs_intermediate}

Having established that different $\bstar$ prescriptions lead to equally good descriptions of low-energy DY data and to compatible TMDs in the non-perturbative region, it is natural to ask how these choices affect the behavior of the extracted distributions at larger transverse momenta.
In this section, we focus on the intermediate- and large-$|\kperp|$ regions of the TMDs obtained from the fits discussed in the previous Section. This aspect of the TMD formalism is often ignored in phenomenological applications, and it has been addressed only recently within alternative frameworks (see Refs.~\cite{Aslan:2024nqg, Simonelli:2025kga}).

The region $|\kperp| \approx Q$ is reliably described by fixed-order perturbative calculations, where the TMD PDF reduces to the OPE expression in Eq.~\eqref{eq:tmd_NLO_kTspace}. Between this limit and the genuinely non-perturbative domain, there exists an intermediate region in which perturbative QCD still applies but fixed-order calculations are insufficient, and large logarithms of the type $\log(Q/|\kperp|)$ must be resummed.
Since this resummation is entirely perturbative, the behavior of the TMDs in this region is expected to be universal and independent of any phenomenological modeling, including the specific prescription adopted to regulate the Landau pole.
However, this expectation is difficult to realize when working in impact-parameter space.
The inverse Fourier transform required to obtain momentum-space distributions inevitably involves large values of $|\bT|$, where prescriptions are introduced and combined with non-perturbative modeling. As a consequence, the behavior of TMDs in the intermediate-$|\kperp|$ region may reflect this nontrivial interplay.

\begin{figure}[h]
\centering
\includegraphics[width=0.6\textwidth]{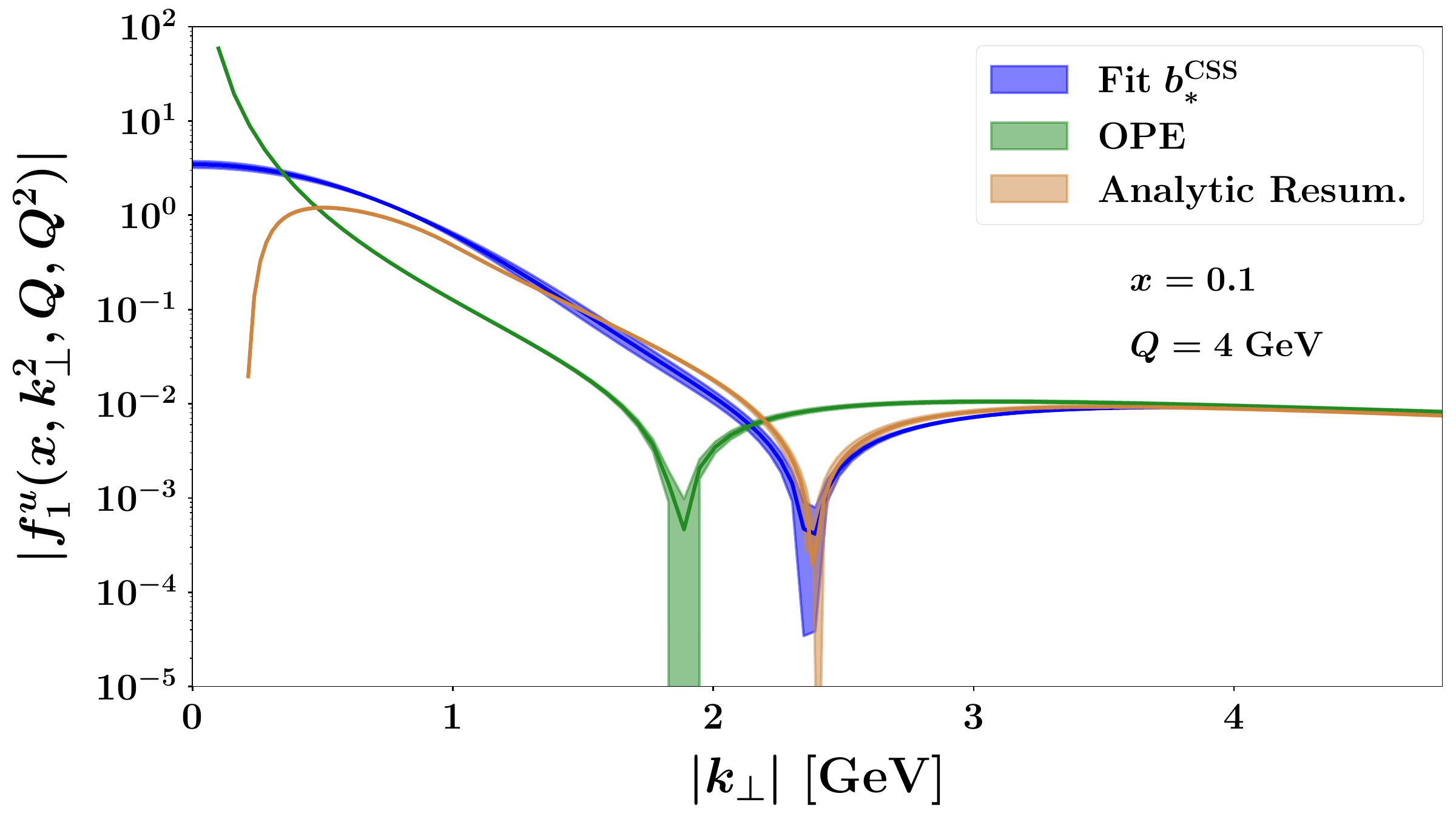}
\caption{Comparison between the TMD PDF of the up quark in a proton at $\mu = \sqrt{\zeta} = Q = 4$ GeV and $x$ = 0.1 as a function of the partonic transverse momentum $|\kperp|$, as extracted from \textit{Fit $\bstar^{\rm CSS}$} (blue curve), the perturbative (fixed order OPE at NLO) result (green curve), and the perturbative resummed (at NLL) result (brown curve). Uncertainty bands represent one-$\sigma$ error.}
\label{f:TMDinter_bstarCSS}
\end{figure}

In Fig.~\ref{f:TMDinter_bstarCSS}, we show the comparison between the unpolarized TMD PDF for the up quark in the proton and $x = 0.1$ at $\mu = \sqrt{\zeta} = Q = 4$ GeV as a function of the quark transverse momentum $|\kperp|$ for the configuration \textit{Fit $\bstar^{\rm CSS}$} (blue band) and the results of perturbative calculations at fixed order (green curve) and resummed directly in transverse momentum space~\cite{Simonelli:2025kga} (brown curve).
The perturbative curves are used as benchmarks to assess the reliability of the extracted TMD in kinematic regions not constrained by data. Ideally, the TMD should match the OPE result at $|\kperp| \approx Q$ and, more generally, follow the resummed prediction for $|\kperp| \gtrsim $1-2 GeV.
We plot the absolute value of the TMD distribution on a logarithmic scale for a better visualization of its features in the intermediate-$|\kperp|$ region. The uncertainty bands represent the one-$\sigma$ error.
First, we observe that the fitted TMD curve approaches the OPE calculation at $|\kperp| \simeq Q$. In particular, the TMD distribution extracted with this configuration of the $\bstar$-prescription closely matches the perturbative OPE tail. We stress that this behavior emerges naturally, without being imposed through the non-perturbative model.
Most importantly, the fit correctly reproduces the resummed behavior, at least within the considered kinematics. This indicates that the region beyond the constraining power of the data is consistently described within this kinematic range.

In this context, a particularly relevant aspect is the relative position of the nodes, \textit{i.e.}, the points where the curves change sign, visible as cusps when the absolute value of the distribution is displayed. This is a specific feature of the transverse momentum dependent operators, as they vanish at $|\bT| = 0$ in conjugate space~\cite{Collins:2016hqq}. The position of these nodes provides interesting information: the resummation of large logarithms, which dominate the intermediate-$|\kperp|$ region, predicts a node at larger transverse momentum compared to fixed-order calculations, and the fitted curve reproduces this behavior.
This suggests that fixed-order expressions become unreliable at transverse momenta of the order of the node scale.

\begin{figure}[h]
\centering
\includegraphics[width=0.48\textwidth]{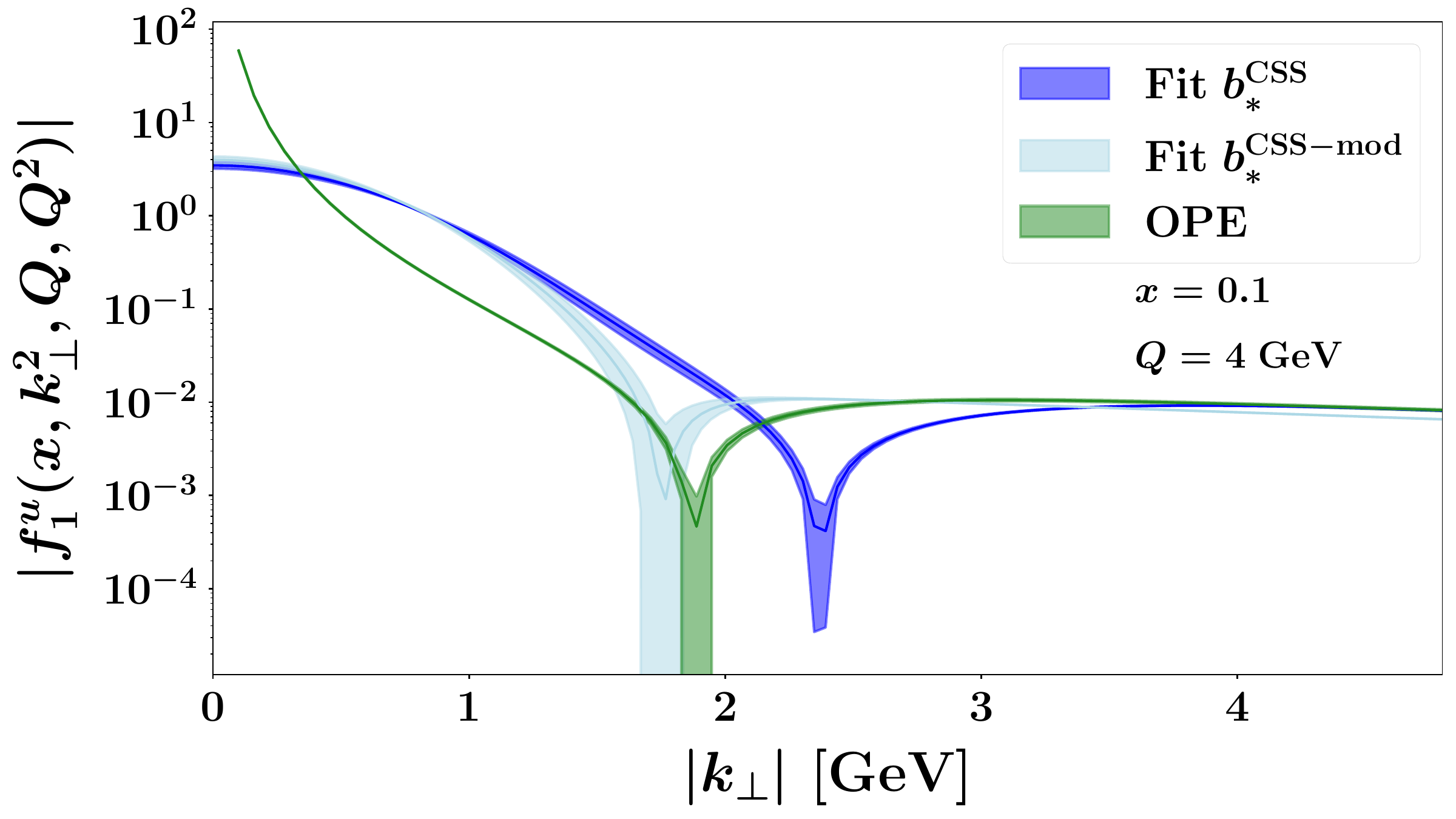}
\includegraphics[width=0.48\textwidth]{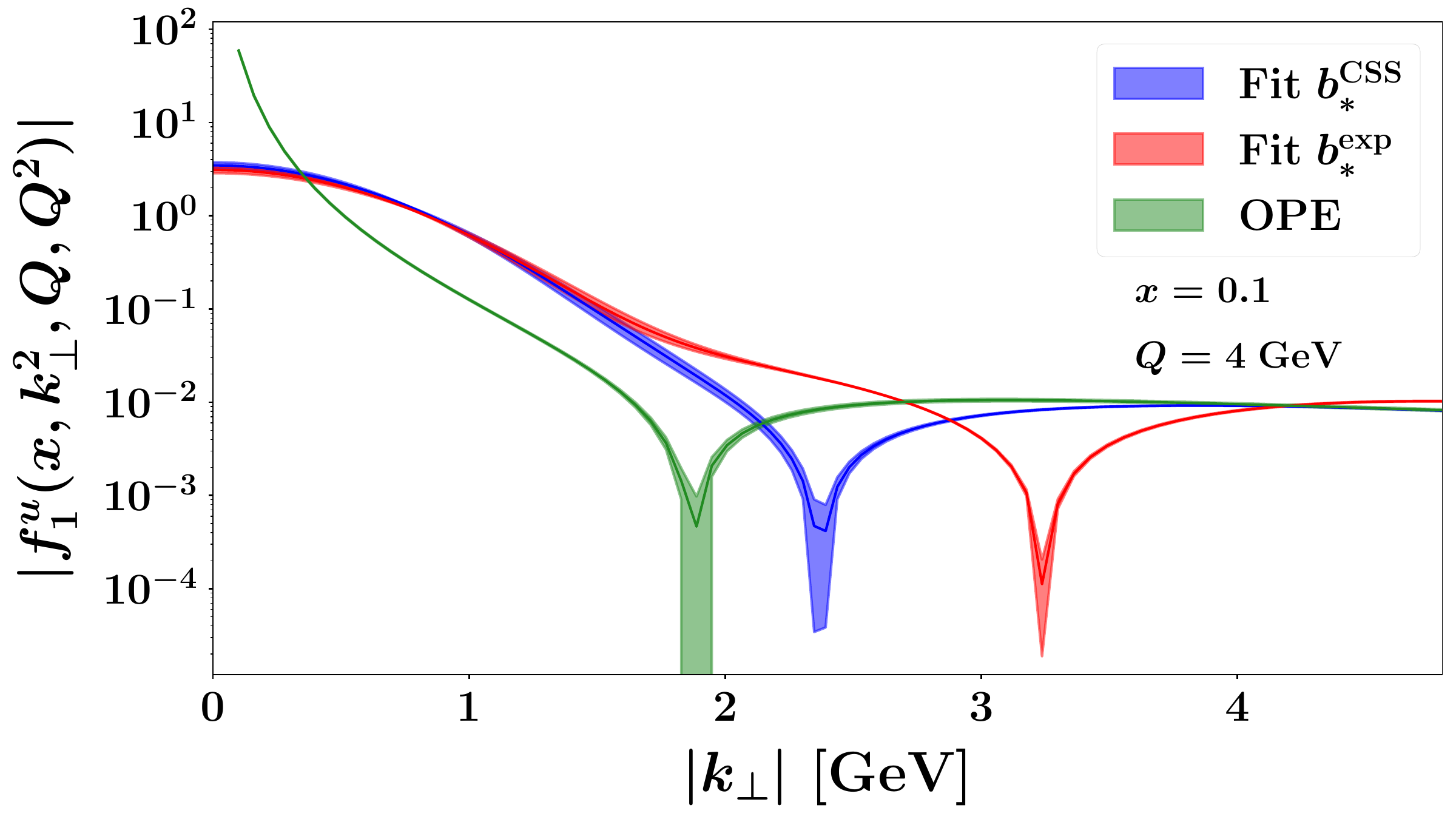}
\caption{Comparison between the TMD PDF of the up quark in a proton at $\mu = \sqrt{\zeta} = Q = 4$  GeV and $x$ = 0.1 as a function of the partonic transverse momentum $|\kperp|$ and the perturbative (OPE) result (green curve). Left panel: comparison between \textit{Fit $\bstar^{\rm CSS}$} (blue band), \textit{Fit $\bstar^{\text{\scriptsize CSS-mod}}$} (light-blue band). Right panel: comparison between \textit{Fit $\bstar^{\rm CSS}$} (blue band), \textit{Fit $\bstar^{\rm exp}$} (red band). Uncertainty bands represent one-$\sigma$ error.}
\label{f:TMDinter_comp}
\end{figure}

The ideal behavior exhibited by \textit{Fit $\bstar^{\rm CSS}$} is not necessarily reproduced by other $\bstar$ configurations. Their impact is examined in Fig.~\ref{f:TMDinter_comp}, where we compare, over the full $|\kperp|$ range, the TMDs extracted in the configurations \textit{Fit $\bstar^{\rm CSS}$} and \textit{Fit $\bstar^{\rm CSS\text{-}mod}$} (left panel), and \textit{Fit $\bstar^{\rm CSS}$} and \textit{Fit $\bstar^{\rm exp}$} (right panel).
In the left panel of Fig.~\ref{f:TMDinter_comp}, we observe that the TMD extracted from \textit{Fit $\bstar^{\text{\scriptsize CSS-mod}}$} (light-blue curve) is consistent with \textit{Fit $\bstar^{\rm CSS}$} (blue curve) in the small- and large-$|\kperp|$ regions, while there are significant differences at intermediate $|\kperp|$.
In particular, when $\bmax = c_1/ 2$ the node of the fitted TMD moves at smaller transverse momentum, even below the fixed-order node, and the large-$|\kperp|$ tail reaches the OPE prediction at smaller $|\kperp|$ than in the $\bmax = c_1$ configuration.
The same behavior is observed when comparing \textit{Fit $\bstar^{\text{\scriptsize exp-mod}}$} to \textit{Fit $\bstar^{\rm exp}$}, indicating that this feature does not strongly depend on the specific $\bstar(\bT^2)$ functional form.
In the right panel of Fig.~\ref{f:TMDinter_comp}, we observe similar features in the comparison of the TMD extracted from \textit{Fit $\bstar^{\rm CSS}$} (blue curve) with the one extracted from \textit{Fit $\bstar^{\rm exp}$} (red curve). Specifically, the two curves are compatible at small $|\kperp|$ and approach the OPE tail at large $|\kperp|$, while they present differences in the intermediate region. The TMD extracted from \textit{Fit $\bstar^{\rm exp}$} agrees with \textit{Fit $\bstar^{\rm CSS}$} up to a larger value of transverse momentum compared to the left panel, with its node occurring at larger $|\kperp|$.

This analysis demonstrates that, within the standard CSS approach, the behavior of TMDs at intermediate transverse momentum is sensitive to the phenomenological choice of the $\bstar$ prescription. This sensitivity can be interpreted as an intrinsic theoretical uncertainty of the method. While the agreement with experimental data remains stable across different $\bstar$ configurations (Sec.~\ref{ss:TMDs_lowEnergy}), a more stringent test could be performed by studying DY processes at higher energies, where data are expected to be more sensitive to the intermediate-$|\kperp|$ region (Sec.~\ref{ss:bstar_HighEnergy}).
Importantly, all results presented here are obtained using the same non-perturbative parametrization (Eq.~\eqref{e:fNP}), implying that discrepancies with perturbative resummation in the intermediate region are artifacts of the approach, due to the unavoidable correlation between the $\bstar$ prescription and the non-perturbative model. In principle, each $\bstar$ implementation should be paired with a dedicated non-perturbative parametrization to reproduce the correct perturbative behavior; in practice, this is unfeasible due to the vast number of possible functional forms. A potential solution could be provided by neural-network representations of the non-perturbative model (see Ref.~\cite{Bacchetta:2025ara}), at the cost of reduced interpretability compared to analytic expressions.

\subsection{Extrapolation to high-energy Drell-Yan data}
\label{ss:bstar_HighEnergy}

A possible way to assess the reliability of a specific choice of $\bstar$–prescription within the standard CSS approach is to test its predictive power at higher energy scales. To this end, we use the TMD distributions extracted from the various fit configurations to compute theoretical predictions for $Z$-boson production at the kinematics of the \textsc{CDF} Run I data~\cite{CDF:1999bpw}.

\begin{figure}[h]
\centering
\includegraphics[width=0.48\textwidth]{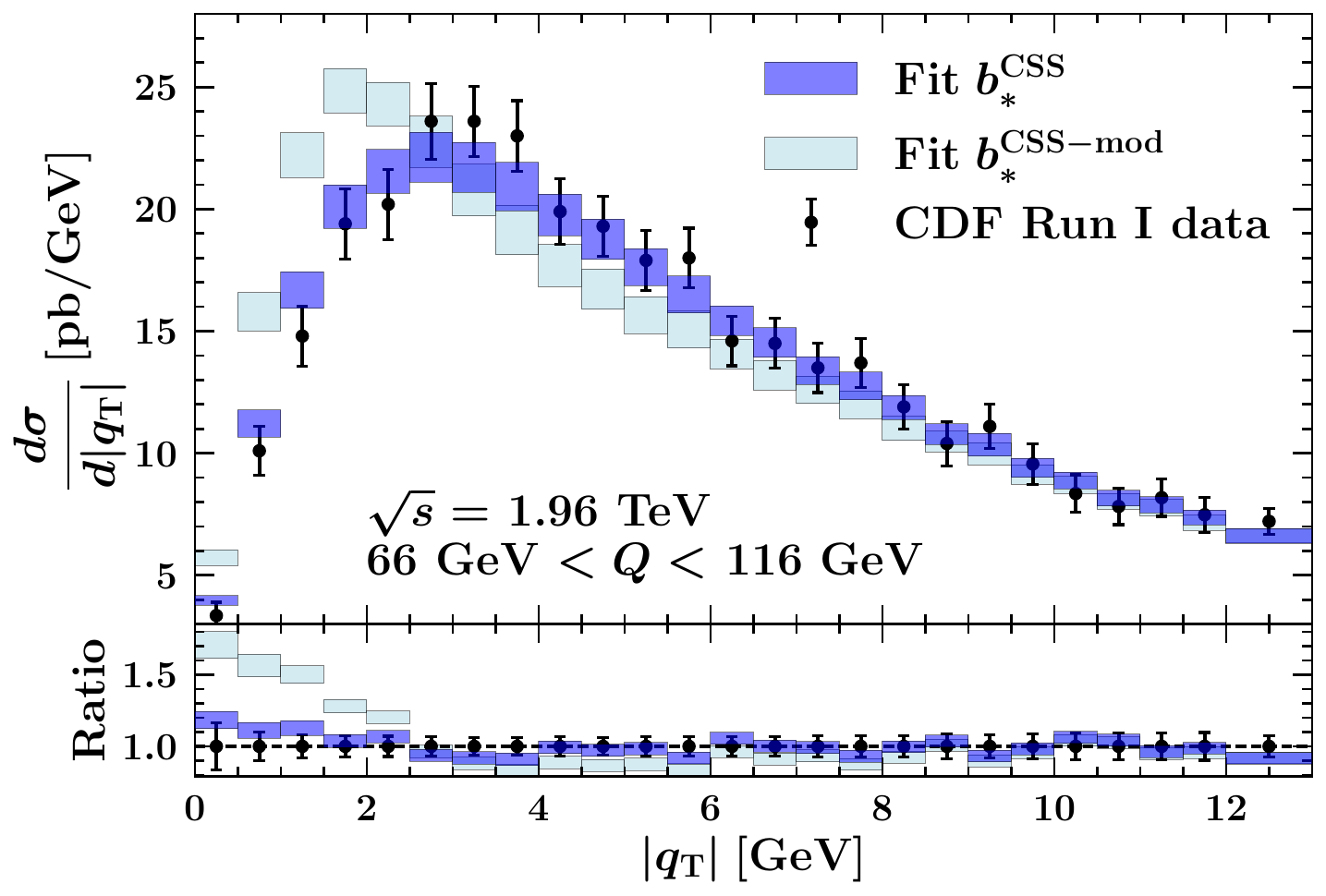}
\includegraphics[width=0.48\textwidth]{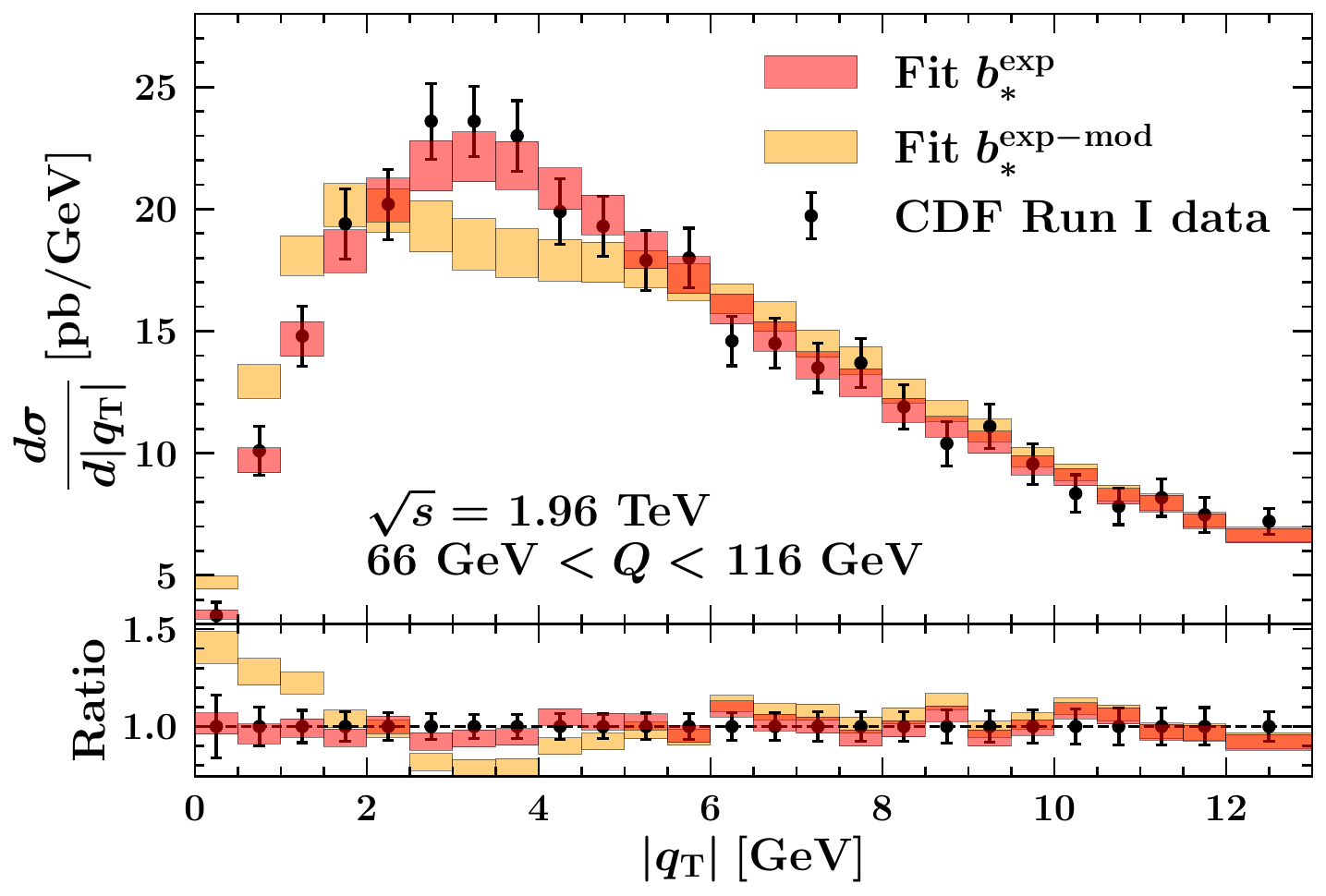}
\caption{Comparison between \textsc{CDF} Run I experimental data and theoretical predictions for $Z$-boson production cross section differential in $|\qT|$. Left panel: comparison between \textit{Fit $\bstar^{\rm CSS}$} (blue curve), \textit{Fit $\bstar^{\text{\scriptsize CSS-mod}}$} (light-blue curve). Right panel: comparison between \textit{Fit $\bstar^{\rm exp}$} (red curve), \textit{Fit $\bstar^{\text{\scriptsize exp-mod}}$} (orange curve).}
\label{f:bstar_CDF}
\end{figure}

In Tab.~\ref{t:chipred_bstar}, we report the breakdown of the central replica $\chi^2$ values normalized to the number of data points ($N_{\text{dat}}$) for the predictions on the considered dataset.
\begin{table}[h!]
\centering
\begin{tabular}{|p{1.5cm}|C{1.5cm}|C{1.5cm}|C{1.5cm}|C{1.5cm}|C{1.5cm}|}
\hline
\multicolumn{2}{|c|}{} & \multicolumn{4}{c|}{\rule{0pt}{2.5ex} $\chi^2 / N_{\text{dat}}$} \\ \hline
Dataset & $\ndat$ & \rule{0pt}{2.5ex} $\bstar^{\rm CSS}$ & $\bstar^{\text{\scriptsize CSS-mod}}$ & $\bstar^{\rm exp}$ & $\bstar^{\text{\scriptsize exp-mod}}$ \\ \hline
\textsc{CDF I} & 25 & 0.83 & 6.16 & 0.57 & 2.87  \\ \hline
\end{tabular}%
\caption{Summary of $\chi^2$ values normalized for the number of data points $\ndat$ of the predictions for \textsc{CDF} Run I data based on various fits performed in this analysis.}
\label{t:chipred_bstar}
\end{table}
Interestingly, we observe that the predictive power of the TMDs extracted from the different fit configurations is not similar and strongly depends on the $\bstar$ configuration.
Figure~\ref{f:bstar_CDF} illustrates the comparison between our predictions (colored bands) and experimental data (black points). The left panel compares \textit{Fit $\bstar^{\rm CSS}$} (blue) with \textit{Fit $\bstar^{\text{\scriptsize CSS-mod}}$} (light-blue), and the right panel compares \textit{Fit $\bstar^{\rm exp}$} (red) with \textit{Fit $\bstar^{\text{\scriptsize exp-mod}}$} (orange). The lower panels show the theory/data ratios.
The \textit{Fit $\bstar^{\rm CSS}$} and \textit{Fit $\bstar^{\rm exp}$} reproduce the experimental data very well, whereas predictions from \textit{Fit $\bstar^{\text{\scriptsize CSS-mod}}$} and \textit{Fit $\bstar^{\text{\scriptsize exp-mod}}$} show a significant discrepancy. This difference can be attributed to the enhanced sensitivity of high-energy data to the intermediate-$|\kperp|$ region of the TMDs, where the extracted distributions show the largest differences (see Fig.~\ref{f:TMDinter_comp}).
In addition, the predictions based on \textit{Fit $\bstar^{\text{\scriptsize CSS-mod}}$} and \textit{Fit $\bstar^{\text{\scriptsize exp-mod}}$} also fail to describe the small-$|\qT|$ region.
This behavior indicates that distortions of the distribution at intermediate $|\qT|$ are compensated by modifications at small $|\qT|$, reinforcing the conclusion that the phenomenological impact of TMDs at intermediate $|\kperp|$ can be significant, especially at high energies.
A legitimate way to address this issue is to include the high-energy data in the fits. In this way, the non-perturbative model may accommodate the tension at low $|\qT|$ observed in Fig.~\ref{f:bstar_CDF}. With this perspective, we perform a new fit including \textsc{E288}, \textsc{E605} \textit{and} \textsc{CDF} datasets. In Tab.~\ref{t:chitable_full}, we report the breakdown of the central $\chi_0^2$ values normalized to the number of data points ($N_{\text{dat}}$) for the various fit configurations.
\begin{table}[h!]
\centering
\begin{tabular}{|p{1.5cm}|C{1.5cm}|C{1.5cm}|C{1.5cm}|C{1.5cm}|C{1.5cm}|}
\hline
\multicolumn{2}{|c|}{} & \multicolumn{4}{c|}{\rule{0pt}{2.5ex} $\chi^2 / N_{\text{dat}}$} \\ \hline
Dataset & $\ndat$ & \rule{0pt}{2.5ex} $\bstar^{\rm CSS}$ & $\bstar^{\text{\scriptsize CSS-mod}}$ & $\bstar^{\rm exp}$ & $\bstar^{\text{\scriptsize exp-mod}}$ \\ \hline
\textsc{E288} & 130 & 0.98 & 0.89 & 1.01 & 0.96 \\ \hline
\textsc{E605} & 50 & 1.72 & 2.12 & 1.38 & 1.96 \\ \hline
\textsc{CDF I} & 25 & 0.74 & 3.38 & 0.56 & 2.19 \\ \hline
Total & 205 & 1.13 & 1.49 & 1.04 & 1.36 \\ \hline
\end{tabular}%
\caption{Summary of $\chi^2$ values and total $\chi^2$ normalized for the number of data points $\ndat$ of the various fit configurations.}
\label{t:chitable_full}
\end{table}
We observe that by including the \textsc{CDF} Run I measurements in the fit, the agreement with these data improves, with the $\chi^2/N_{\text{dat}}$ decreasing from 6.16 to 3.38 for \textit{Fit $\bstar^{\text{\scriptsize CSS-mod}}$} and from 2.87 to 2.19 for \textit{Fit $\bstar^{\text{\scriptsize exp-mod}}$}, although a sizable tension remains.
However, this improvement leads to a deterioration in the description of the \textsc{E605} dataset, reflecting the fact that the fit is forced to accommodate competing constraints across different energy regimes.

This demonstrates that, for the specific non-perturbative model, the \textit{Fit $\bstar^{\text{\scriptsize CSS-mod}}$} and \textit{Fit $\bstar^{\text{\scriptsize exp-mod}}$} configurations lead to TMDs that develop features at intermediate transverse momentum incompatible with perturbative resummation (see Fig.~\ref{f:TMDinter_bstarCSS}). As a result, these fits fail to provide a reliable description of high-energy data.
In other words, TMDs consistent with the result of analytic resummation yield satisfactory predictions at higher scales. 
Notably, even \textit{Fit $\bstar^{\text{\scriptsize exp}}$}, despite departing from the perturbative result, successfully describes high-energy measurements.
This suggests that, in absence of a perturbative benchmark, simultaneous fits to low- and high-energy data may extract multiple, equally valid TMD distributions at intermediate partonic transverse momentum.
Global fits take advantage of the complementary constraining power of low-energy data on the small-intrinsic-momentum region of TMD distributions and of high-energy data on the intermediate-momentum region.
As a result, within the standard approach, such global analyses are necessary for a reliable extraction of TMDs, although some residual phenomenological bias may remain.
Finally, we emphasize that the inclusion of high-energy data remains essential, as it typically probes different kinematic regions in $x$, thereby providing crucial information on the $x$-dependence of TMD PDFs.

\subsection{Impact of small-$b_T$ region}
\label{ss:smallbT}

In CSS formalism, the region at low $|\bT|$ values of TMDs is described by fixed-order perturbative QCD and therefore is expected to have limited influence on cross sections at low transverse momentum, which are dominated by non-perturbative physics. However, an examination of this region expose apparent inconsistencies that must be properly addressed in order to preserve a consistent definition of TMDs. 

A first concern regards the connection between TMDs and (collinear) parton distributions. Naively, one might expect that the latter can be obtained simply by integrating the transverse-momentum distribution over all values of $\kperp$
\begin{align}
\label{eq:naive_integral}
    &f_{q/h}(x,\mu) \stackrel{?}{=} \int d\vec{k}_T  F_{q/h}\big(x,|\kperp|,\mu, \zeta/\mu^2\big) \equiv
    \fourier{F}_{q/h}\big(x,0,\mu, \zeta/\mu^2\big) \, ,
\end{align}
that formally coincides with the TMD in Fourier space evaluated at $|\bT|=0$.
However, resummation predicts that TMDs vanish as a positive power of $|\bT|$ in the limit $|\bT| \to 0$~\cite{Collins:2016hqq}.
This implies that the integral in Eq.\eqref{eq:naive_integral} is zero, failing to reproduce the corresponding collinear PDFs. 
The origin of this mismatch lies in the fact that the transverse momentum integral is ultraviolet (UV) divergent and thus cannot be defined without introducing an appropriate regulator to correctly address the region $|\kperp| \gg \mu$ (see, \textit{e.g.}, Refs.~\cite{Collins:2011zzd,Collins:2021vke,Ebert:2022cku}).
For collinear PDFs, standard perturbative schemes such as $\overline{\text{MS}}$ are sufficient to yield well-defined results. However, when they are reconstructed from intrinsically non-perturbative TMDs, these schemes are inadequate and alternative prescriptions are required~\cite{Ebert:2022cku}.
A widely-used solution is to shift the problem from the integral to the integrand. In this interpretation, the predicted vanishing of TMDs at small $|\bT|$ is regarded as an artifact of extrapolating the resummation beyond its domain of validity, in the UV region. To circumvent this, the resummation variable $|\bT|$ is then modified \emph{ad hoc} in the problematic region. 
In practice, TMDs are regulated by introducing an additional short-distance scale $\bmin = c_1/\mu$, ensuring that they saturate at $\bmin$ as $|\bT| \to 0$. 
This interpretation simply provides a practical way to regularize the UV divergence of the integral by acting directly on the integrand, with $\bmin$ functioning as an explicit UV regulator.
A minimal implementation consists in replacing the logarithms $\log{(\mu/\mu_b)}$ in the Sudakov factor with $\log{(1 + |\bT|/\bmin)}$, ensuring regular behavior in the limit $|\bT| \to 0$ (see, \textit{e.g.}, Refs.~\cite{Parisi:1979se,Bozzi:2005wk}).
Most often $\bmin$ is implemented as an additional constraint to the standard approach, complementing the saturation of the perturbative content at $\bmax$~\cite{Boer:2014tka,Collins:2016hqq,Bacchetta:2017gcc}.
All implementations, however, modify the low-$|\bT|$ behavior of the TMDs, even in the region $|\bT| \sim c_1/\mu$, where fixed-order perturbative expansions are expected to provide reliable approximations of the operators. As a consequence, the matching with the OPE at large $|\kperp|$ observed in Fig.~\ref{f:TMDinter_bstarCSS} can be significantly spoiled. 

A further motivation sometimes put forward in favor of the inclusion of $\bmin$ concerns the behavior of cross sections at low transverse momentum. 
Starting at NLL accuracy, analytic resummation performed directly in transverse-momentum space develops spurious divergences, which can compromise the stability of theoretical predictions~\cite{Frixione:1998dw}. 
These singularities originate from the small-$|\bT|$ region and can be traced back to a mismatch between the small value of $\qT^2$ and the potentially large values of its two-dimensional components entering the Fourier transform.
The introduction of the scale $\bmin$ automatically regulates this problematic region.
However, it has been shown in Ref.~\cite{Simonelli:2025kga} that a careful treatment of the small-$|\bT|$ domain when performing the inverse Fourier transform leads to well-defined cross sections. This can be achieved without introducing the additional scale $\bmin$, though at the price of a more involved analytic structure. 
Therefore, there is no fundamental physical requirement to incorporate $\bmin$ in the formalism. 
The issue ultimately reduces to a trade-off: on the one hand, $\bmin$ restores the integral (unitarity) relation between TMDs and collinear PDFs; on the other, it compromises the natural matching between the small- and large-$q_T$ regimes. Its adoption is therefore a phenomenological choice, balancing theoretical consistency against practical simplicity.

\bigskip

\begin{figure}[h]
\centering
\includegraphics[width=1.\textwidth]{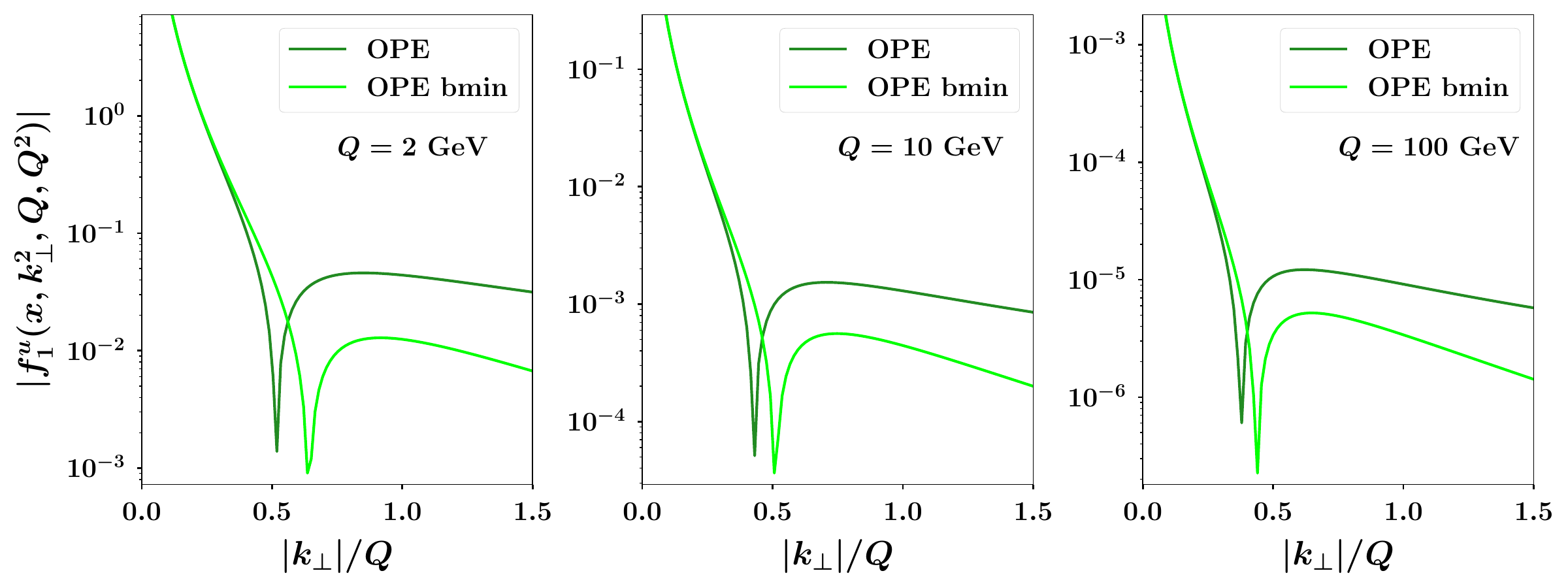}
\caption{Comparison between the perturbative OPE result (green curve), and the OPE result including $\bmin$ (light-green curve) for the absolute value of the $u$-quark TMD PDF at $x=0.1$ and $Q=2, 10, 100$ GeV (from left to right).}
\label{f:all_ope}
\end{figure}

As a concrete example, we consider the $\bstar$ implementation as in Eq.~\eqref{e:bstar_collins} and we apply the modification proposed in Ref.~\cite{Collins:2016hqq}:
\begin{equation}
\label{e:bstar_collins_bmin}
\bar{\bb}_*^{\rm CSS}(\bT^2) = \sqrt{\frac{\bT^2 + \bmin^2}{1 + \frac{\bT^2+\bmin^2}{\bmax^2}}} \, ,
\end{equation}
At small $|\bT|$, the parametrization above reduces to $\sqrt{\bT^2 + \bmin^2}$. This modification propagates to the OPE of Eq.~\eqref{eq:tmd_NLO_kTspace}, which now reads~\cite{Rogers:2024cci}:
\begin{align}
\label{eq:tmd_NLO_kTspace_bmin}
    &F_{q/h}(x,|\kperp|,\mu,\zeta/\mu^2;\bmin) = 
    \frac{a_S(\mu)}{2\pi |\kperp|^2}
    \int_x^1 \frac{d \hat{x}}{\hat{x}}
\notag \\
&\quad
    4 C_F \bigg \{
    \bmin |\kperp|\,K_1\big(\bmin |\kperp|\big) 
     \frac{(1 + \hat{x}^2)}{(1-\hat{x})_+} +
     \notag \\
     &\quad+
     \bigg[
     K_0\big(\bmin |\kperp|\big)
     +
     \bmin |\kperp|\,
     K_1\big(\bmin |\kperp|\big) 
     \bigg(
     \log{\Big(\bmin \frac{|\kperp|}{c_1}\Big)}
     +
     \log{\frac{\zeta}{\kperp^2}}
     \bigg)
     \bigg]
     \delta(1-\hat{x})
   \bigg \} \, f_{q/h} \Big( \frac{x}{\hat{x}};\mu \Big)
   \notag \\
   &\quad+
   2\,\bmin |\kperp|\,K_1\big(\bmin |\kperp|\big)
   \Big[\hat{x}^2 + (1-\hat{x})^2 \Big]\,f_{g/h} \Big( \frac{x}{\hat{x}};\mu \Big)
    +
   \mathcal{O}(a_S^2)  \, ,
\end{align}
where $K_0$ and $K_1$ are modified Bessel function of the second kind. The expression above reduces to Eq.~\eqref{eq:tmd_NLO_kTspace} in the limit $\bmin \to 0$, however, it represents a \emph{different} OPE. In particular, the matching at $|\kperp| \sim Q$ is significantly distorted for any value of $Q$, despite the expectation that the effect of $\bmin = c_1/\mu$ should become negligible at higher energies\footnote{Note that the limit $|\kperp| \to Q$ has to be taken \emph{before} the limit $\bmin \to 0$.}. 
\begin{figure}[h]
\centering
\includegraphics[width=0.6\textwidth]{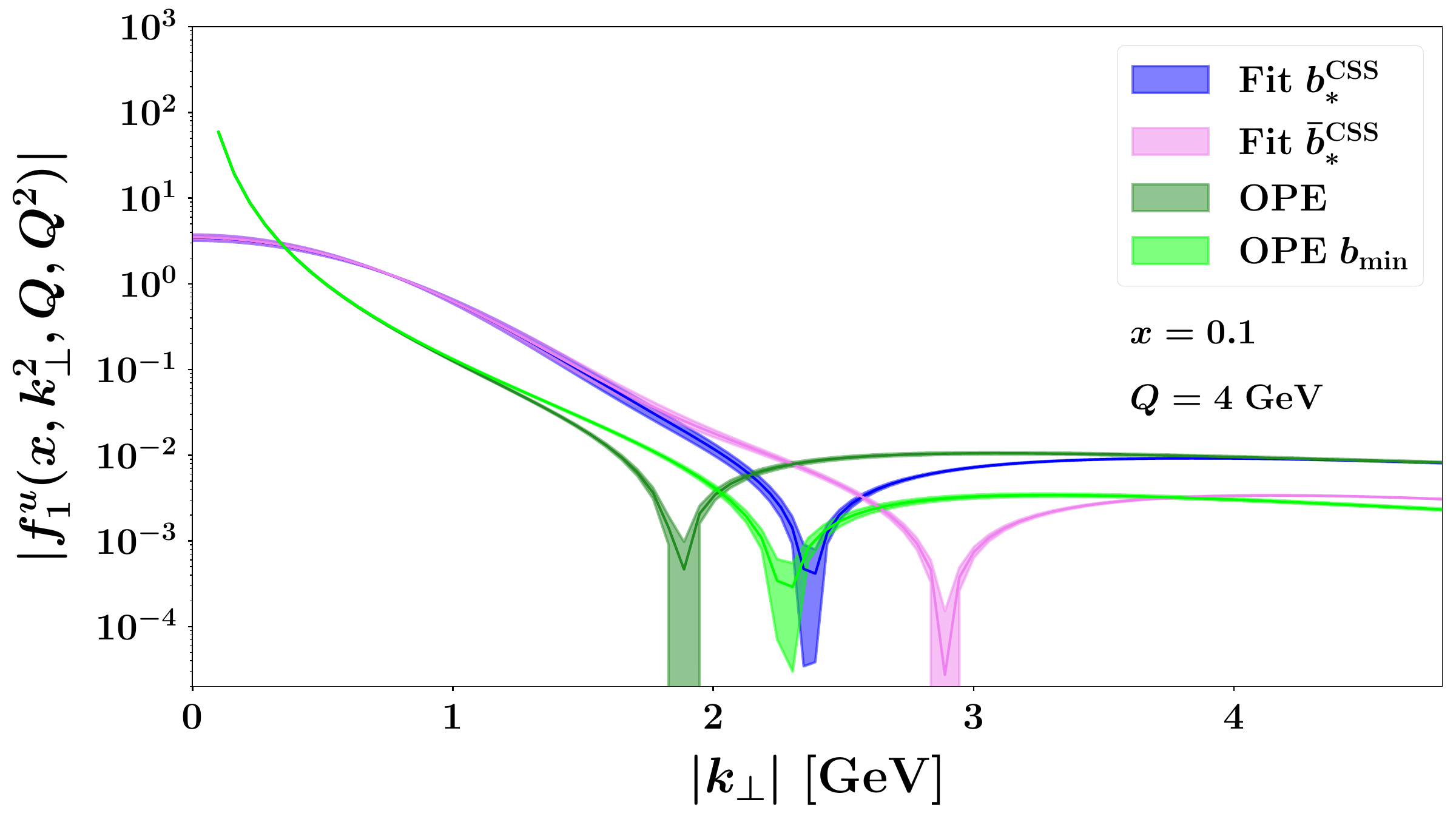}
\caption{Comparison of the TMD PDF of the up quark in a proton at $x=0.1$ and $\mu = \sqrt{\zeta} = Q = 4$~GeV, shown as a function of the partonic transverse momentum $|\kperp|$. Results are given for the \textit{Fit $\bb_*^{\rm CSS}$} (blue band), the \textit{Fit $\bar{\bb}_*^{\rm CSS}$} (light-violet band), the perturbative OPE result (green band), and the OPE result including $\bmin$ (light-green band). The uncertainty bands correspond to the 1$\sigma$ error.}
\label{f:TMD_bstarmin}
\end{figure}
We illustrate this distortion in Fig.~\ref{f:TMD_bstarmin}, where we display the extracted TMDs from the \textit{Fit $\bar{\bb}_*^{\rm CSS}$} (light-violet band), namely the counterpart of the \textit{Fit $\bstar^{\rm CSS}$} (blue band) in which the TMDs are supplemented with the $\bmin$ prescription of Eq.~\eqref{e:bstar_collins_bmin}.
We compare them with the fixed-order perturbative OPE including $\bmin$ (light-green curve) and with the standard OPE result (green curve). The deviation of the two curves is manifest in the figure: the large-$|\kperp|$ behavior of the OPE including $\bmin$ is significantly different from the standard OPE
\footnote{This discrepancy is not resolved by increasing the perturbative accuracy (see discussion in Ref.~\cite{Simonelli:2025kga}).}.
In Fig.~\ref{f:all_ope}, we compare the OPE calculation with and without the introduction of $\bmin$ (light-green and green curves, respectively) at $x=0.1$ and at different values of the energy scale $Q$. We immediately see that the two results are significantly different at large transverse momentum even at very large energy scale.

Despite the distortions observed in the TMD distributions at large transverse momentum, the impact of $\bmin$ appears to be negligible for TMD phenomenology. This is indicated by the consistency of the blue and light-violet band at low-$|\kperp|$ in Fig.~\ref{f:TMD_bstarmin}. 
Indeed, repeating the fit presented in the previous sections with the inclusion of $\bmin$ leads to a similar description of the experimental data.
Note that the modification of Eq.~\eqref{e:bstar_echevarria} is implemented following the MAP extractions (see, \textit{e.g.}, Refs.~\cite{Bacchetta:2017gcc,Bacchetta:2019sam,Bacchetta:2022awv}).
Therefore, we conclude that if phenomenological analyses are restricted to the low-transverse-momentum region, the introduction of the additional scale $\bmin$ corresponds to a specific phenomenological choice, with a small impact on the predictions.
By contrast, analyses aiming to describe the full transverse-momentum spectrum of a given observable should take care of the sizable discrepancies observed at the level of the TMD distributions. 
These effects may require the introduction of additional prescriptions (such as ad hoc profile functions) to ensure a controlled matching to the fixed-order collinear-factorization framework.
Such a strategy would constitute another additional phenomenological ingredient, potentially challenging the natural matching between different kinematic regimes.

The concrete benefit associated with the introduction of $\bmin$ is in the integration over the entire $|\kperp|$ spectrum. With $\bmin=c_1/\mu$, Eq.~\ref{eq:naive_integral} becomes: 
\begin{align}
\label{eq:bmin_integral}
    &\int d\vec{k}_T  F_{q/h}\big(x,|\kperp|,\mu, 1;\bmin\big) 
    = 
    f_{q/h}(x,\mu) + a_S(\mu) \big[\Delta^{[1]}_{q/j} \otimes f_{j/h}\big] (x,\mu) + \mathcal{O}(a_S(\mu)^2)
\end{align}
where the coefficients $\Delta_{j/j'}$ coincide with the non-logarithmic part of the Wilson coefficients of the OPE (see Eq.~\eqref{eq:tmd_NLO_kTspace}). 
This procedure thus provides a controlled connection between TMDs and collinear PDFs.
It should be noted that this equivalence may be subject to corrections of order $\mathcal{O}(\bmin/\bmax)$, depending on the specific implementation of the $\bmin$ prescription. In particular, this issue is present in the implementation of Eq.~\eqref{e:bstar_collins_bmin}, but does not appear in the prescription that generalizes Eq.~\eqref{e:bstar_echevarria}.
Despite its practical advantages, we stress that the introduction of $\bmin$ is not a prerequisite for defining the transverse-momentum integral, but rather one possible choice for its ultraviolet regularization. 
Another legitimate and theoretically sound alternative consists in defining the integral with an explicit ultraviolet cutoff, \textit{i.e.} restricting the integration to $|\kperp| \leq \mu$. 
This scheme was also proposed in Ref.~\cite{Gonzalez-Hernandez:2022ifv} as an alternative to the $\overline{\text{MS}}$ prescription for standard PDFs.
Its main advantage is that it avoids modifying the TMDs, thereby preserving the matching to the fixed-order result at large $|\qT|$.
However, this approach complicates the connection to collinear PDFs. In this case, the integral receives contributions from the intrinsic non-perturbative component of the TMD:
\begin{align}
\label{eq:cutoff_integral}
    &\int^{\mu} d\vec{k}_T  F_{q/h}\big(x,|\kperp|,\mu, 1\big) 
    = 
    f_{q/h}(x,\mu) + a_S(\mu) \big[\Delta^{[1]}_{q/j} \otimes f_{j/h}\big] (x,\mu) + \mathcal{O}(a_S(\mu)^2) + \mathcal{O}\Big(\frac{1}{\mu \bmax}\Big)
\end{align}
where the suppressed term reflects the residual sensitivity to the non-perturbative region. Such contributions roughly scale as $1/(\mu \bmax)$ and become increasingly relevant as the scale $\mu$ is lowered.
A possible estimate of the size of this non-perturbative corrections can be obtained by comparing the TMDs  extracted in the fit of the previous Sections with the full perturbative result of Eq.\eqref{eq:bmin_integral}.
\begin{figure}[h]
\centering
\includegraphics[width=0.49\textwidth]{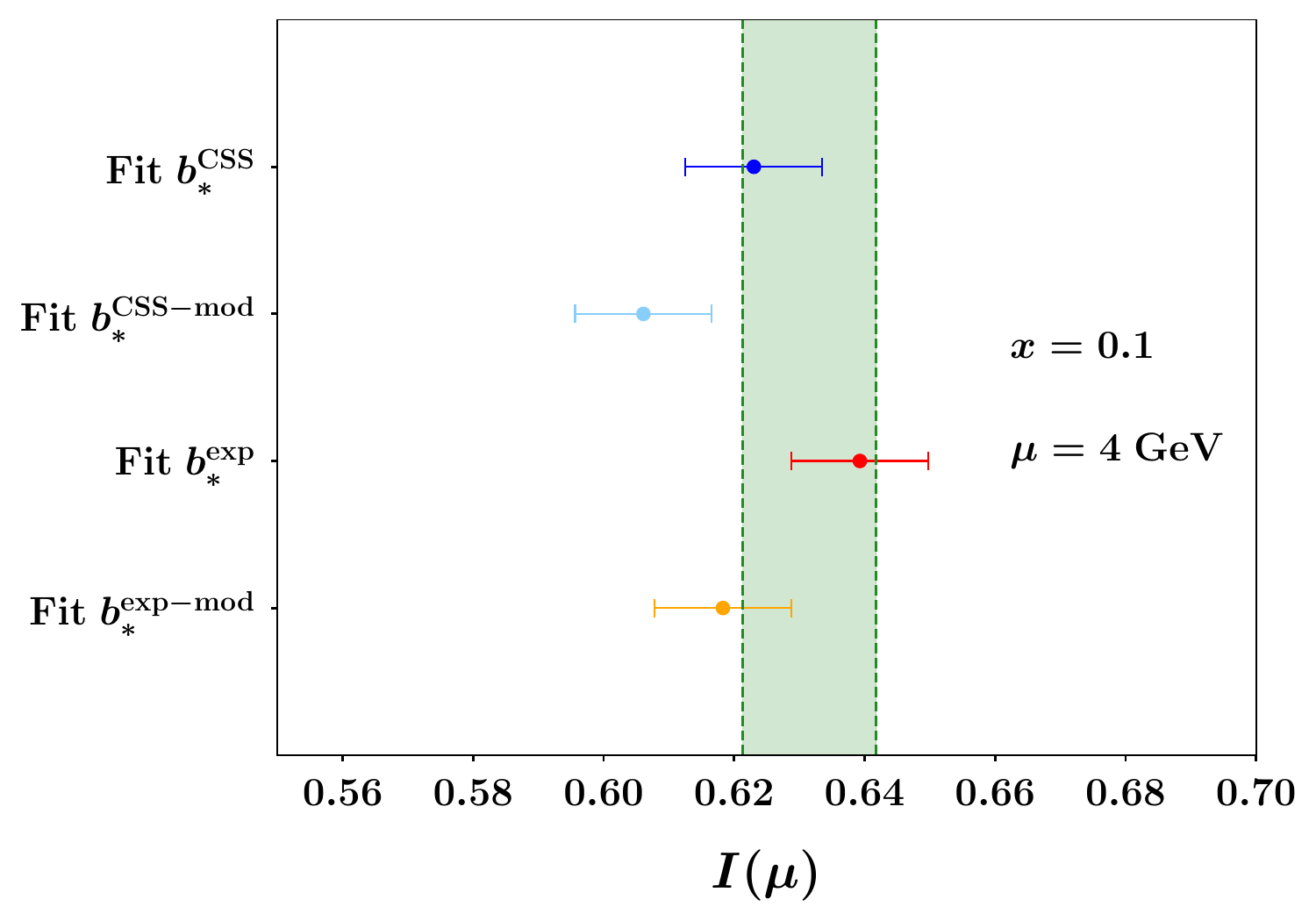}
\includegraphics[width=0.49\textwidth]{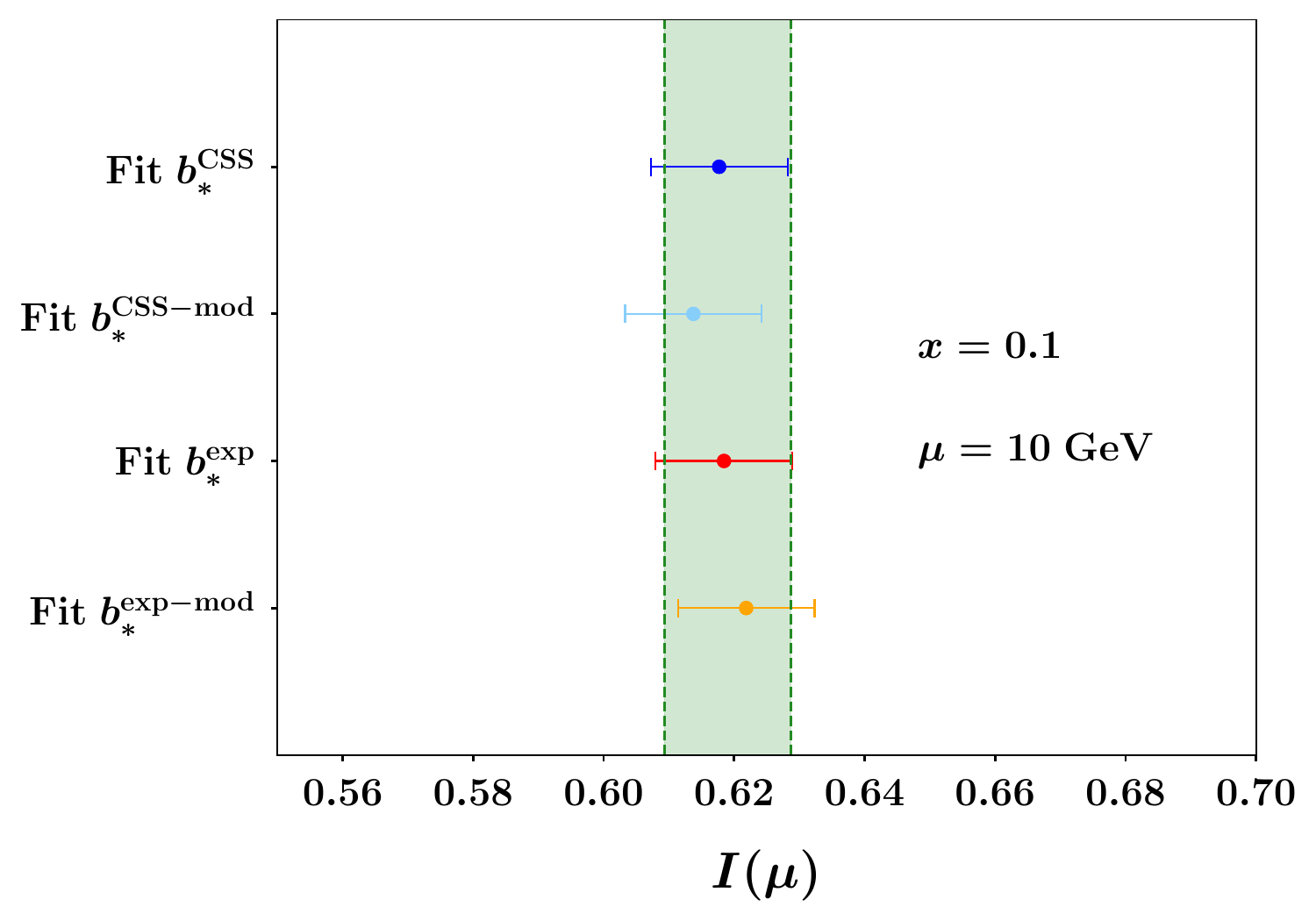}
\caption{Integral of the TMD PDF of the up quark in a proton at $\mu = \sqrt{\zeta} = Q = 4$ (left panel) and $10$ (right panel)  GeV and $x$ = 0.1 as extracted from \textit{Fit $\bstar^{\rm CSS}$} (blue point), \textit{Fit $\bstar^{\rm exp}$} (red point), \textit{Fit $\bstar^{\text{\scriptsize CSS-mod}}$} (light-blue point), and \textit{Fit $\bstar^{\text{\scriptsize exp-mod}}$} (orange point) and collinear PDF result (green band). The uncertainty bands represent the 68\% C.L.}
\label{f:bstar_integrals}
\end{figure}
To this end, we define
\begin{align}
    \label{eq:IQ_def}
    &I(\mu) = 
    \int^{\mu} d\vec{k}_T  F_{q/h}\big(x,|\kperp|,\mu, 1\big) \,.
\end{align}
In Fig.~\ref{f:bstar_integrals}, we display $I(\mu)$ for the various fit configurations at $x = 0.1$ and for two different values of the scale $\mu$, set to $4$ GeV (left panel) and $10$ GeV (right panel), and compare it with the collinear PDF (green band) corrected by the NLO factor on the r.h.s of Eq.\eqref{eq:cutoff_integral}.
We observe that, with the only exception of \textit{Fit $\bstar^{\text{\scriptsize CSS-mod}}$} (light-blue point), the TMD integrals are compatible with the collinear PDF within one standard deviation. This demonstrates that TMDs extracted within the standard CSS framework, even without the inclusion of $\bmin$, satisfy the integral relation discussed above to good accuracy, provided that the integral itself is UV regularized.
A similar conclusion was reached in Ref.~\cite{delRio:2024vvq}. 
We also confirm that the agreement between collinear PDFs and regularized TMD integrals improves as the scale increases, indicating that the non-perturbative corrections in Eq.\eqref{eq:cutoff_integral} are indeed suppressed at higher values of the scale, as can be inferred by comparing the right panel and the left panel of Fig.~\ref{f:bstar_integrals}.

\bigskip

\section{Conclusions}
\label{s:conclusions}

In this work we have presented a detailed phenomenological study of transverse-momentum dependent distributions within the Collins-Soper-Sterman (CSS) framework. 
This work has focused on clarifying the role of various prescriptions used to treat the interface between parturbative and nonperturbative regimes, and on assessing their effect on the extraction and interpretation of TMDs.
Our analysis has focused in particular on the interplay between perturbative resummation, phenomenological modeling, and experimental constraints across different kinematic regimes.

Within the CSS formalism, resummation is carried out in impact-parameter space where it controls the large logarithms arising at large transverse separations $|\bT|$. 
To regularize the behavior of TMD distributions near the Landau cut a $\bstar$ prescription is often introduced. This procedure smoothly freezes the impact parameter at a scale $\bmax$ and therefore necessitates the inclusion of suitable non-perturbative multiplicative factors.
In the formulation of the CSS framework, the arbitrariness associated with the choice of prescription is expected to be compensated by the parametrization of the non-perturbative contributions. 
However, as we have discussed, such compensation is not realized in practice. 
As a result, the extracted TMD distributions retain a dependence on the specific functional form of the $\bstar$ prescription and on the associated parameter choices.

In this work, we have examined how these prescriptions are implemented in a realistic phenomenological setting and we have quantified their impact through a systematic comparison of different $\bstar$ functional forms and choices of $\bmax$.
All configurations considered in our study provide an equally good description of low-energy Drell-Yan data, and lead to TMD distributions that are compatible in the low-transverse-momentum region. 
At large transverse momentum, the extracted TMDs behave as expected, progressively approaching the perturbative result, although at different rates depending on the chosen configuration. 
In contrast, clear differences emerge in the intermediate transverse-momentum region, particularly where the TMD becomes negative. In this regime, some configurations reproduce remarkably well the behavior obtained from analytic resummation in transverse-momentum space, suggesting that they provide a more reasonable representation of the underlying dynamics.
These differences do not, however, correspond to a systematic uncertainty in the extracted TMDs when confronted with high-energy data. By studying predictions for high-energy observables, such as the CDF $Z$-boson transverse-momentum spectrum, we have shown that configurations with modified values of $\bmax$ fail to reproduce the experimental measurements. 
Repeating the fit with the inclusion of high-energy data does not change this conclusion. 
Notably, the $\bstar$ configurations that are more consistent with analytic resummation are also those that yield good agreement with experimental data.
This demonstrates one of the main results of our analysis: within the standard CSS approach, performing global fits that combine low- and high-energy data is essential. 
Low-energy measurements are primarily sensitive to the non-perturbative core of the TMDs, while high-energy data probe the effects of evolution and, consequently, the prescription that governs
the interface between perturbative and non-perturbative physics.

In other words, global fits mitigate the limited control over the impact of the prescriptions by simultaneously constraining different kinematic regions, but they do not remove completely the underlying scheme dependence.
The use of highly sophisticated parametrizations, such as neural networks, can be regarded as the most general realization of this strategy where the non-perturbative modeling is sufficiently flexible to absorb any residual arbitrariness. However, such implementations shift the emphasis from specifying a physically motivated functional form to constraining a probability band spanning multiple admissible shapes. The physical interpretation of the extracted TMDs then becomes less direct.
In this sense, a possible alternative is to develop frameworks that differ from the standard CSS implementation and do not rely on prescriptions as external technical tools, but instead incorporate them as intrinsic components of the non-perturbative modeling. These alternative strategies are relatively recent and still require further exploration. They typically sacrifice maximal flexibility for a more constrained and physically transparent structure, but the analytic complexity is increased, demanding for a higher phenomenological effort.

We have also examined the behavior of TMDs in the small-$|\bT|$ region and its implications for the relation between the TMD integral and collinear parton distributions. In particular, we examine the $\bmin$ prescription within the broader framework of ultraviolet renormalization schemes for integrated quantities.
Since this prescription acts at the level of the integrand in the small-$|\bT|$ region, it modifies the perturbative structure of the TMD, affecting the region where fixed-order expansions apply.
However, we have verified that its impact at low $|\qT|$ is numerically small. For applications focused on the non-perturbative intrinsic part of TMDs, including $\bmin$ does not significantly affect the results and preserves a clear connection with integrated distributions and observables.
We have further compared this prescription with alternative schemes, in particular with a cutoff approach in which the integral is restricted to $|\kperp| \leq \mu$. This procedure introduces power-suppressed corrections, which we have shown to be numerically small and to decrease with the energy scale. 

Overall, this study highlights the importance of carefully assessing the role of phenomenological prescriptions in TMD extractions. While such aspects are often treated as black boxes in phenomenological analyses, we have shown that they can induce systematic effects that influence both the extracted distributions and their interpretation. 
This will be particularly important in view of the next generation of TMD extractions and forthcoming experimental programs at facilities such as the HL-LHC, AMBER, JLab~12, the Electron--Ion Collider, and possible future experiments at SoLID and JLab~22~GeV.

\section*{Acknowledgements}
We gratefully thank Simone Rodini for his valuable comments in the final stages of the manuscript.
A.S. was supported by the Italian Ministry of University and Research (MUR) grant PRIN 2022SNA23K funded by the European Union -- Next Generation EU, Mission 4, Component 2, CUP I53D23001410006.

\bibliographystyle{myrevtex} 
\bibliography{paper}

\end{document}